\documentclass{article}
\usepackage[utf8]{inputenc}
\usepackage{graphicx}
\usepackage{tabularx}
\usepackage[ruled,vlined]{algorithm2e}
\usepackage[english]{babel}
\usepackage{amsthm}
\usepackage{amssymb}
\usepackage{dutchcal}
\usepackage[ruled,vlined]{algorithm2e}
\usepackage{cancel}
\usepackage{amsmath}
\usepackage{subcaption}
\usepackage{booktabs}
\newcommand*{\medcap}{\mathbin{\scalebox{1.5}{\ensuremath{\cap}}}}%

\usepackage{biblatex}
\begin{document}
\title{Hosting Capacity Approach}
\author{Sicheng~Gong, Vladimir~Ćuk, \\Tiago~Castelo~de~Oliveira, J.F.G.~Cobben}
\maketitle

\begin{abstract}
This chapter proposes an evolved concept of "hosting capacity" using the term of "feasible region". Through converting the grid model into a more compact one, "hosting capacity region" not only is promising to further exploit the grid potential for power delivery, but also benefits grid operation feasibility investigation with concise formulas. The hosting capacity region assessment schemes are exploited as well. Facing the derived hosting capacity, originally complicated energy storage optimization problems can be represented algebraically, which is more efficient and friendly for computer processing. Case studies based on a 10.5kV Dutch grid have been implemented, eventually confirming the validity of relevant assessment and optimization methods. \\ (\textit{This work is a preprint of a book chapter. If accepted, the copy of record will be available at the IET Digital Library.})
\end{abstract}

\section{Concept of Hosting Capacity}
\begin{figure}
    \centering
    \includegraphics[width=\linewidth]{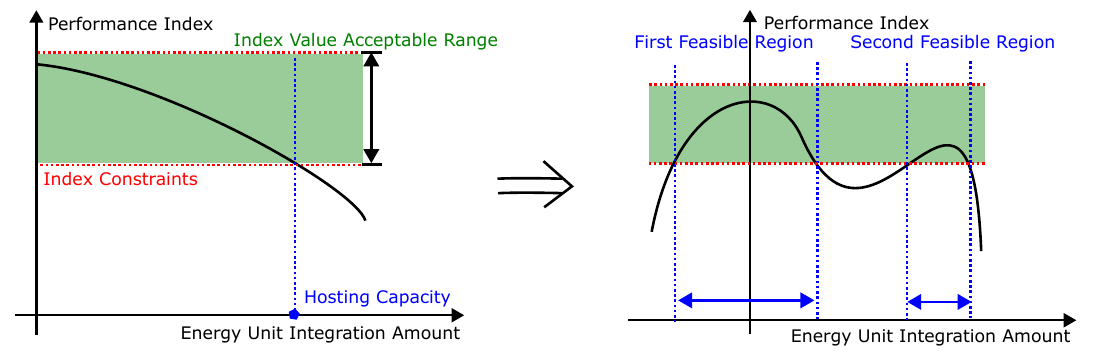}
    \caption{Conceptual evolution of hosting capacity}
    \label{fig:bollen_Concept}
\end{figure}

Referring to ~\cite{bollenintegration}, "hosting capacity" is a term to quantify the acceptable capacity of loads or generations in certain profiles, for integration to specific points of connection (POCs) in an existing grid. It is an essential index in industrial practices to help distributive system operator (DSO) allocate a suitable integration capacity for a energy unit, which is in the waiting list for future grid integration. As illustrated in Fig.~\ref{fig:bollen_Concept}, originally designed for distributive generations or loads, "hosting capacity" is treated as a cut-off value according to some performance index thresholds.

Meanwhile, not all grid performance indices can keep monotonic to the unit integration amount, as quadratic, cubic and even biquadratic equations are ubiquitous when mathematically describing the grid. Moreover, with a rising penetration rate of intermittent energy unit, energy storage system (ESS) and prosumers are expected for grid regulation, which can play roles of generation and load simultaneously, asking for a span over the whole range in horizontal axis. A single cut-off value can no lounger explicitly describe capacity constraints in these scenarios, naturally stimulating the conceptual evolution of "hosting capacity".

As illustrated in Fig.~\ref{fig:bollen_Concept}, facing a non-monotonic vertical-axis-crossing plot, "hosting capacity" can be represented by two disjoint segments. Therefore, intuitively, we take the "feasible region" term as a suitable tool to boost the conceptual evolution of "hosting capacity". Considering more POCs and performance indices investigated, the essence of "hosting capacity" seems more apparent for us, which should be seen as a feasible region of integration capacity. 

\subsection{Performance Index Clarification and Simplification}
In advance of detailed discussions on how such concept evolves, we need turn to its original format, figuring out respective performance indices that deserve our attentions.
Technically, hosting capacity determination should involve in voltage level, cable (or transformer) capacity, harmonic distortion, voltage dip, and etc. These indices seem complicated and mutually influenced, while we can simply categorize them into two aspects: power quality and facility limit. The former one is defined by grid codes and the latter one refers to facility datasheets. 

For the convenience to further exploit the essence of hosting capacity, we mainly investigate the impacts of voltage level and cable capacity on hosting capacity. Such simplification makes sense as this book is talking about energy storage, while harmonic distortions and voltage dips will involve in filters and controllers, eventually overextending our scope. Moreover, instead of considering stochastic unit profiles or unstable cable parameters, a restrained concept of "hosting capacity" is more rational, which is derived based on a deterministic scenario. 

\subsection{Hosting Capacity Region Definition}

So far, we still have a basic question: how to represent the so-called "hosting capacity", through several scalar value thresholds or combinational constraining cuts? Actually, even for only one POC, its integration power can be divided into two aspects: active power and reactive power, and there definitively exists mutual relationship between their thresholds. With a raised number of involved POCs, which is denoted as $n$, the hosting capacity needs to be represented by $2n$ variables, including each POC's integrated active and reactive power. Therefore, we rephrase "hosting capacity" as "hosting capacity region", which is defined to be the combinational feasible region of these $2n$ variables. Definitively, in practices, we can use less dimensions if left variables are assumed constant.

This evolved concept can bring several benefits for DSO, not only extending the grid operation state space, but also converting the grid model into a more computer-friendly one. More illustrations are provided as follows.

\subsubsection{Grid Power Delivery Potential Exploitation}
\begin{figure}
    \centering
    \includegraphics[width=0.95 \linewidth]{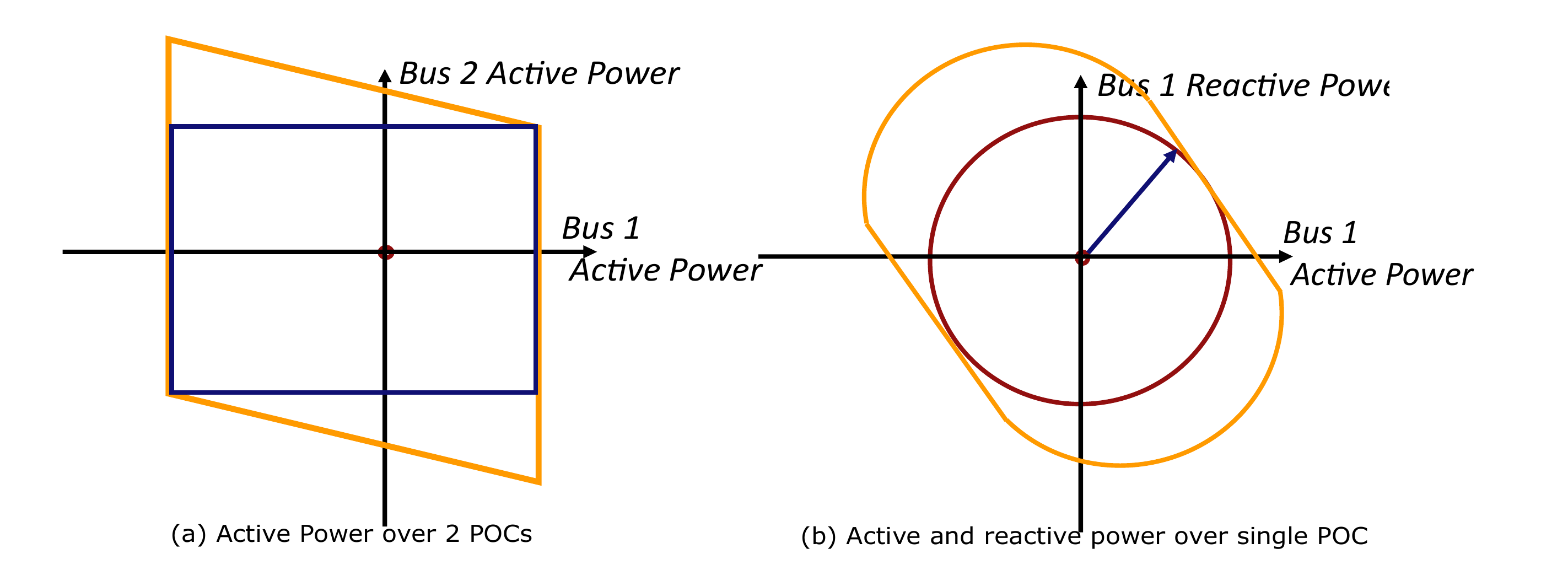}
    \caption{Hosting capacity region cases }
    \label{fig:Capacity_Region_Benefit}
\end{figure}

Starting from the 2-dimensional (2D) perspective, we can look into a combinational feasible region of active power unit integration over two POCs. As shown in Fig.~\ref{fig:Capacity_Region_Benefit}(a), the orange parallelogram denotes their respective feasible region boundary, which is derived in later Section.~\ref{subsubsec:case_2D_linear}. Here we use the results in advance to illustrate how evolved "hosting capacity" exploits the grid potential to deliver more power.

Based on the previous "hosting capacity" concept, the DSO will set separate upper and lower thresholds to both bus operators, and the corresponding feasible region will become a rectangular denoted by blue lines. It can be seen that some state space is dropped accordingly. With Bus~1 injecting more active power, Bus~2 can absorb more active power without violating any voltage or current limits. The original concept covers up such possiblity. Instead, The combinational "hosting capacity region" will cover all possible scenarios, allowing the grid to deliver more power in some extreme scenarios.

In Fig.~\ref{fig:Capacity_Region_Benefit}(b), a similar phenomenon can be observed when focusing on a single POC. By using the results at later Section.~\ref{subsubsec:no_convex_case} in advance, the combinational feasible region are denoted by through the orange boundary. Still using the original concept, the DSO will set a scalar value constrain on apparent power over such POC, leading to a round feasible region. Part of the grid power delivery capacity is wasted eventually. Both cases in Fig.~\ref{fig:Capacity_Region_Benefit} provide a vivid illustration on how such new "hosting capacity region" exploits the grid power delivery potential.

\subsubsection{Grid Problem Format Transformation}
In conventional grid research works, the grid model itself commonly suffers a high data volume, as it needs to include nominal voltage/capacity value, cable/transformer parameters, grid topology and certain energy unit profiles. It is computer-unfriendly for model processing. Even for a simple operation scenario, the computer has to run iterative power flow calculation codes and check relevant constraints. When we use such model for further investigations, frequent power flow calculation and quantities of operation scenarios will cause a high hurdle to power engineers to realize quick and efficient solving.

Meanwhile, through hosting capacity region, the grid model can be saved when we investigate operation feasibility problems, for instance minimal energy storage size. Since such region can be easily processed by advanced computational algebra technology, we successfully transform the previous complicated problems into more compact ones. Moreover, the hosting capacity region contains no enough information to retrieve grid parameters and existing loads. The DSO can share this region to relevant POC operators, encouraging their combinational operation while without extra grid and user data privacy concerns. 

\subsection{Hosting Capacity Region with ESS}
Referring to a derived hosting capacity region, the DSO can set combinational connection capacity limits to POCs accordingly. Meanwhile, in practices, some users are willing to legitimately request a larger connection capacity even with higher connection expenses. The DSO cannot fudge such request. Instead of physically upgrading grid cables or transformers, energy storage equipment installation is a promising alternative to satisfy such request, especially facing high power intermittent energy units, for instance emerging ultra-fast electric-vehicle~(EV) charging infrastructure systems in urban areas. 

\begin{figure}
    \centering
    \includegraphics[width=0.9\linewidth]{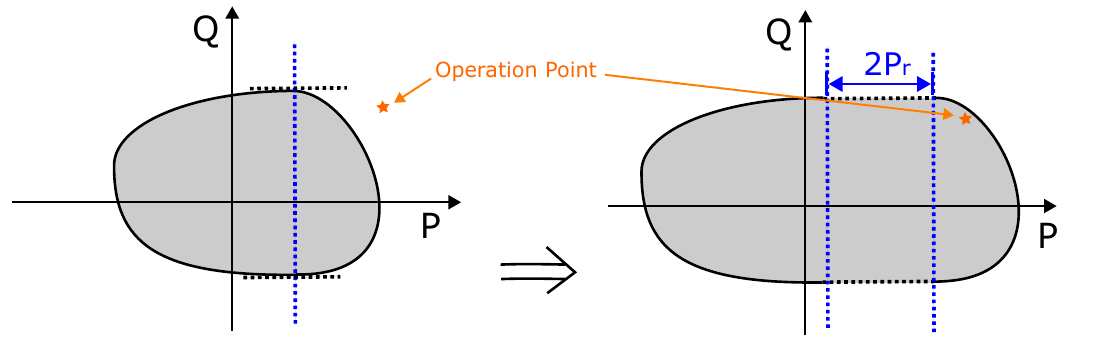}
    \caption{Hosting capacity region reshaping with ESS }
    \label{fig:Capacity_Region_Reshape_ESS_2D}
\end{figure}

Geometrically, with ESS integrated, the relevant hosting capacity region is equivalently reshaped. As illustrated by Fig.~\ref{fig:Capacity_Region_Reshape_ESS_2D}, still looking into a 2D hosting capacity region, with installing bidirectional active power regulation equipment (maximum regulation power is denoted as $P_r$) over the same POC, the region is reshaped accordingly. The operation point is covered by the updated hosting capacity region, indicating such operation status becomes acceptable after ESS integration. The derived minimal value of $P_r$ will be a solid reference for DSO when deploying relevant ESS.

In later parts of this chapter, we will discuss on how to assess the hosting capacity region, and how ESS explicitly imposes impacts on such region. It can be seen from results that the evolved concept of "hosting capacity" will further benefit DSO to optimally allocate grid integration and regulation capacity. 

\section{Hosting Capacity Region Assessment}
\subsection{Distribution Grid Model}

\begin{figure}
    \centering
    \includegraphics[width=0.8\linewidth]{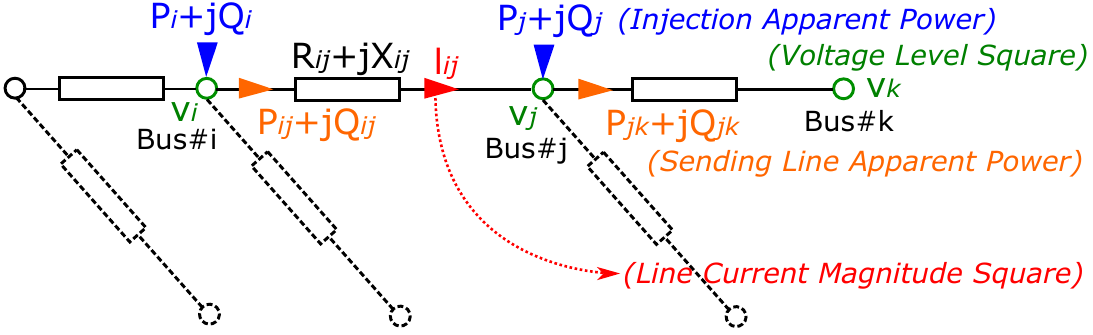}
    \caption{Distribution grid schematic}
    \label{fig:grid_schematic}
\end{figure}

In this book, distribution grids are mainly focused on, which commonly are radial as shown in Fig.~\ref{fig:grid_schematic}. According to \cite{baran1989network}, we use DistFlow model to mathematically describe the distribution grid model as given in (\ref{eq:tight_model}). Corresponding variable definitions in (\ref{eq:tight_model}) are provided in Table~\ref{tab:TableOfNotationForMyResearch}. For further illustration, the coupling between bus injection power and line transmission power is denoted by (\ref{eq:tight_model_1})-(\ref{eq:tight_model_2}). The relationship among bus voltage level, line current level and line transmission power are written in (\ref{eq:tight_model_3})-(\ref{eq:tight_model_4}). Equation~(\ref{eq:tight_model_5})-(\ref{eq:tight_model_8}) are variable constraints defined by unit profiles, grid codes and cable datasheets. 

\begin{subequations}
\begin{align}
    \label{eq:tight_model_1}
    P_{ij}&=\sum_{k: (j,k)\in E } P_{jk}+r_{ij}l_{ij}-P_j \\
    \label{eq:tight_model_2}
    Q_{ij}&=\sum_{k: (j,k)\in E } Q_{jk}+x_{ij}l_{ij}-Q_j \\
    \label{eq:tight_model_3}
    v_j &=v_i-2(r_{ij}P_{ij}+x_{ij}Q_{ij})+(r_{ij}^2+x_{ij}^2)l_{ij}\\
    \label{eq:tight_model_4}
    l_{ij}& = \frac{P_{ij}^2+Q_{ij}^2}{v_i}\\
    \label{eq:tight_model_5}
    P_i^{min}&\leq P_i \leq P_i^{max}\\
    \label{eq:tight_model_6}
    Q_i^{min}&\leq Q_i \leq Q_i^{max}\\
    \label{eq:tight_model_7}
    v_{i}^{min}&\leq v_i \leq v_{i}^{max}\\
    \label{eq:tight_model_8}
    &~~~~ l_{ij} \leq l_{ij}^{max}
\end{align}
\label{eq:tight_model}
\end{subequations}

\begin{table}[htbp]\caption{General notations in DistFlow model}
\centering % to have the caption near the table
\begin{tabular}{c c}
\toprule
Index & Meaning\\
\midrule
\multicolumn{2}{c}{\underline{Constant Parameters}}\\
\multicolumn{2}{c}{}\\
$E$ / $(j,k)$ & Grid graph / Connection between node~j and node~k\\
$r_{ij}$ / $x_{ij}$ & Line resistance / reactance between node~i and node~j\\
$P_i^{min}$ / $P_i^{max}$ & Minimum / Maximum of active power in node~i \\
$Q_i^{min}$ / $Q_i^{max}$ & Minimum / Maximum of reactive power in node~i \\  
$v_i^{min}$ / $v_i^{max}$ & Minimum / Maximum of voltage level square on node~i \\ 
$l_{ij}^{max}$ & Maximum of current magnitude square from  i to j\\ 
\multicolumn{2}{c}{}\\
\multicolumn{2}{c}{\underline{Decision Variables}}\\
\multicolumn{2}{c}{}\\
$P_j$ &  Equivalent injection active power in node~j\\
$Q_j$ &  Equivalent injection reactive power in node~j\\
$v_j$ &  Square of voltage level over node~j\\
$l_{ij}$ &  Square of current magnitude from node~i to node~j\\
$P_{ij}$ &  Injected active power from node~i to node~j\\
$Q_{ij}$ &  Injected reactive power from node~i to node~j\\
\bottomrule
\end{tabular}
\label{tab:TableOfNotationForMyResearch}
\end{table}

Although there are a quantity of decision variables to be solved in a deterministic operation scenario, we only care about the feasibility region of $P_i$ and $Q_i$, since only these variables are concerned when deriving the hosting capacity region. In advance of introducing technical details for hosting capacity region assessment, preliminary analysis on DistFlow model will be implemented. 

However, you may still own a basic question, regarding the cable capacitance ignorance in DistFlow model. There are two ways to reconsider cable capacitance when $\Pi$-model is adopted. The first one is to figure out an equivalent constant reactive power load over each POC, as each POC voltage level is constrained and we can ignore such deviations. The second one is to create more POCs with zero voltage level and without any energy unit integrated. Those POCs are connected to original POC with pure capacitive lines. It will definitely increase our computation burden when deriving the hosting capacity region, while no accuracy sacrifice is caused.

\subsubsection{Ergodic Testing}
Instead of mathematical processing for hosting capacity region derivation, with high-performance computers, we can still get a rough view through ergodic testing. Although an unconstrained quantity of testing scenarios may not be very elegant for an engineer to answer this question, they still deserve in the initial stage to help find some clues.

Normally, when discussing about hosting capacity, $P_i$ and $Q_i$ are assumed determined, then numerical methods are adopted to solve (\ref{eq:tight_model_1}-\ref{eq:tight_model_6}). Through checking whether $l_{ij}$ and $v_i$ meet (\ref{eq:tight_model_7}-\ref{eq:tight_model_8}), the respective operation point will be evaluated. However, in this part, we try to look into the same problem reversely, and implement ergodic testing with $v_i$ and $l_{ij}$ varying. With known $v_i$ and $l_{ij}$, $P_i$ and $Q_i$ can be calculated directly as shown in (\ref{eq:ergoric_model_VI}), where power flow calculation can be directly saved. Through solving quadratic equations, we can derive several feasible operation points at one time.

Meanwhile, even keeping the same sampling resolution on $v_i$ and $l_{ij}$, the total testing scenario number will increase exponentially with grid scale raising. In some practices, it is unnecessary for us to derive the combinational hosting capacity region of all POCs in the grid. Therefore, instead of (\ref{eq:ergoric_model_VI}), we can keep the conventional method of ergodic $P_j$ and $Q_j$ testing, especially when the expected hosting capacity region is in less dimensions.

\begin{subequations}
\begin{align}
    v_j &=v_i-2(r_{ij}P_{ij}+x_{ij}Q_{ij})+(r_{ij}^2+x_{ij}^2)l_{ij}\\
    \Longrightarrow Q_{ij}&=aP_{ij}+b,~a\gets \frac{-r_{ij}}{x_{ij}},~b\gets\frac{v_i-v_j+(r_{ij}^2+x_{ij}^2)l_{ij}}{2x_{ij}}\\
     \Longrightarrow P_{ij}^2&+(aP_{ij}+b)^2-v_i l_{ij}=0~(\text{considering}~ l_{ij}= \frac{P_{ij}^2+Q_{ij}^2}{v_i})\\
     \Longrightarrow P_{ij}&,~Q_{ij}~\text{are solved by root formula}\\
    \Longrightarrow P_{j}&\gets\sum_{k: (j,k)\in E } P_{jk}+r_{ij}l_{ij}-P_{ij}~,~Q_{j}\gets\sum_{k: (j,k)\in E } Q_{jk}+x_{ij}l_{ij}-Q_{ij} 
\end{align}
\label{eq:ergoric_model_VI}
\end{subequations}

\subsubsection{Non-convex Hosting Capacity Region}
\label{subsubsec:no_convex_case}
\begin{figure}
    \centering
    \includegraphics[width=0.7\linewidth]{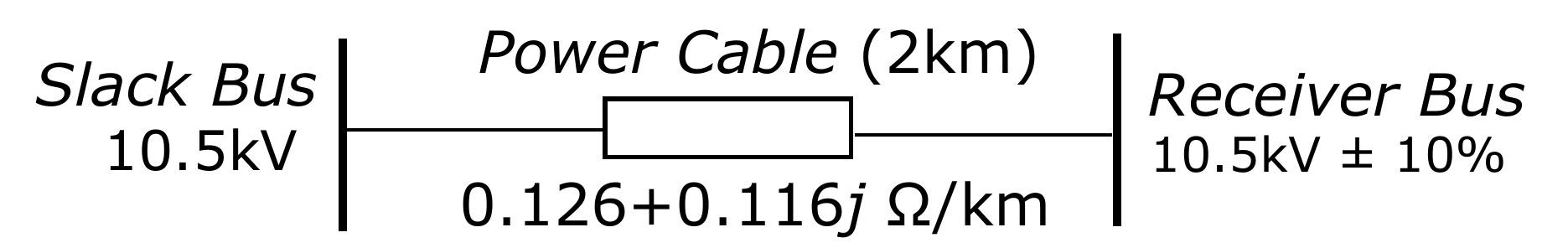}
    \caption{Schematic of a simple grid case }
    \label{fig:Simple_Case_Schematic}
\end{figure}

The introduced ergodic testing method can be employed in a simple grid case as shown in Fig.~\ref{fig:Simple_Case_Schematic}. Such grid is composed of a slack bus, a receiver bus and a piece of power cable. According to grid codes, the receiver bus owns its voltage level flexibility of 10\%. 

\begin{figure}
     \centering
     \begin{subfigure}[b]{0.47\linewidth}
         \centering
         \includegraphics[width=\textwidth]{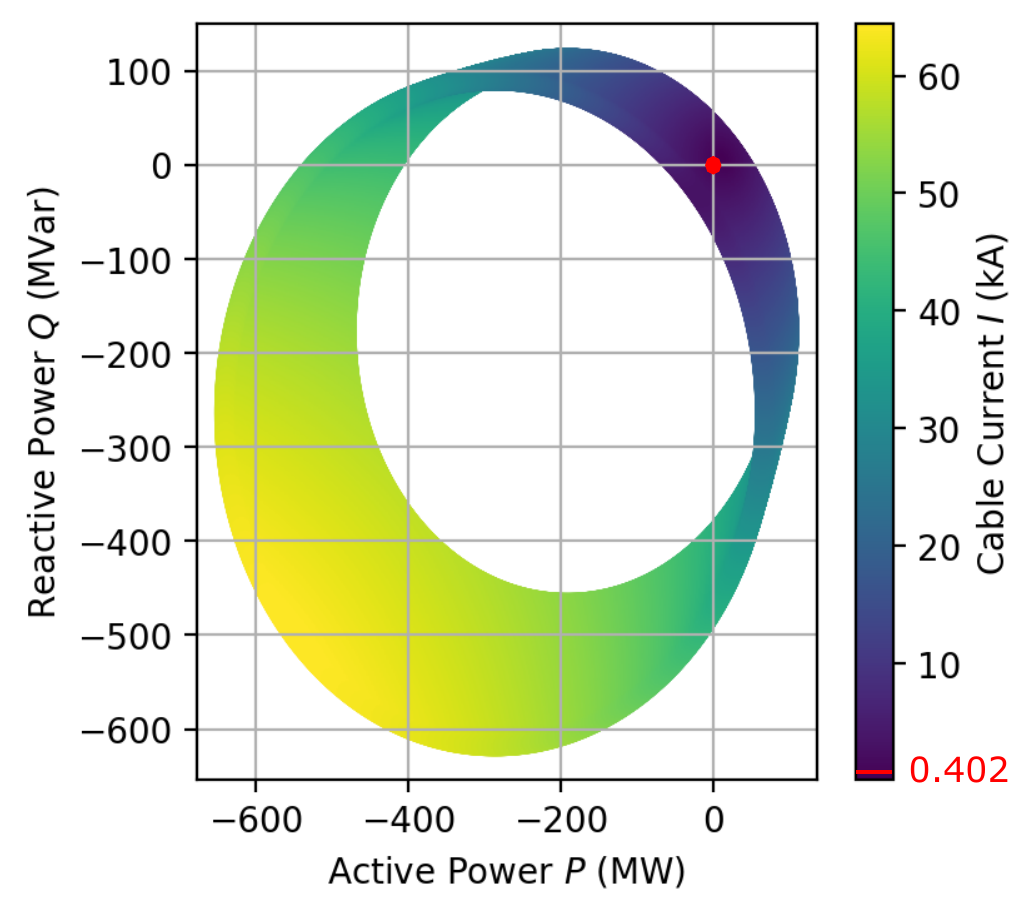}
         \caption{Overall Region }
         \label{fig:Simple_Case_Region_Overall}
     \end{subfigure}
\begin{subfigure}[b]{0.47\linewidth}
         \centering
         \includegraphics[width=\textwidth]{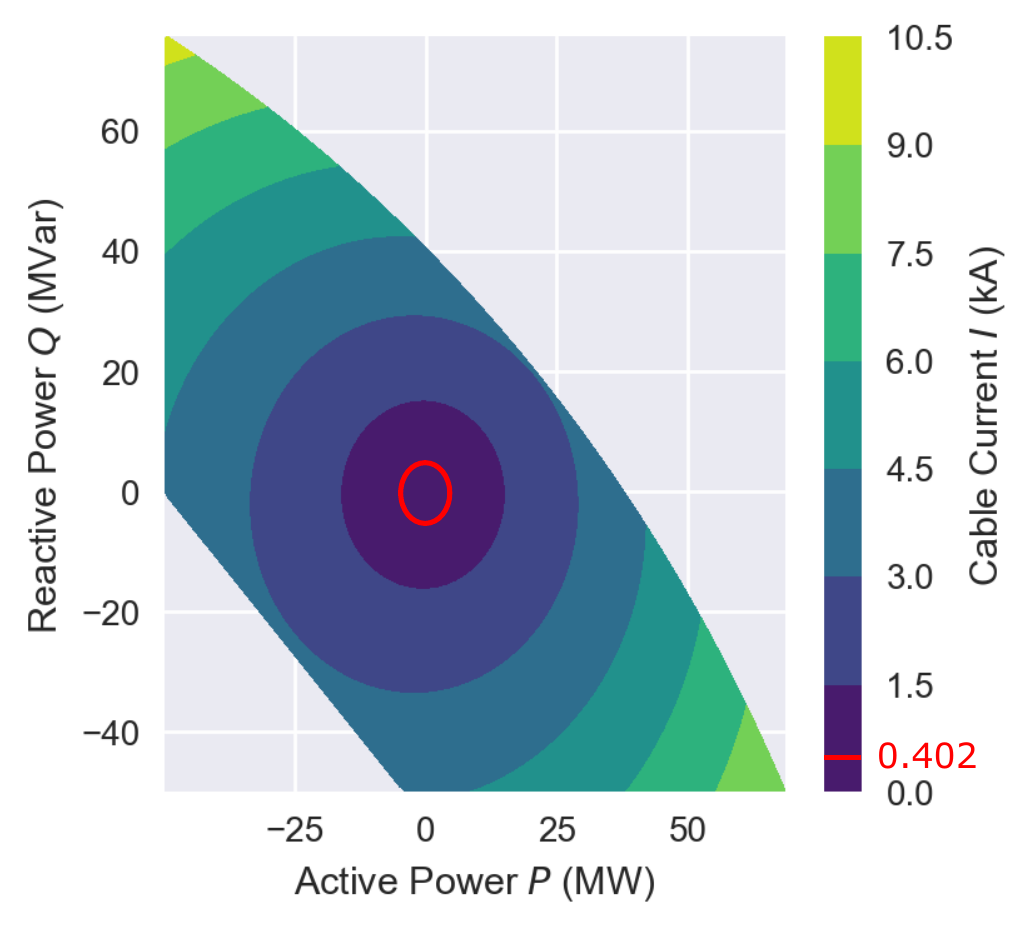}
         \caption{Zoomed Region }
         \label{fig:Simple_Case_Region_Contour}
     \end{subfigure}
        \caption{Hosting Capacity Region of a simple grid case}
        \label{fig:Simple_Case_Region}
\end{figure}

Without power cable current limits, taking the receiver bus as a POC, its respective hosting capacity region is given in Fig.~\ref{fig:Simple_Case_Region_Overall}, where horizontal and vertical axis indicate active and reactive power separately. The colorbar illustrates corresponding cable current level distribution in such region. Geometrically, the hosting capacity region is non-convex. That sounds not a good news, as non-convexity will commonly cause extra obstacles when we try to numerically solve some geometrical problems. Fortunately, current level constraints are inevitable in industrial practices. According to cable datasheet, it current level threshold is equal to 402A. We highlight practical feasible region by red contours in Fig.~\ref{fig:Simple_Case_Region_Overall}, which is seen as a small red dot.

Moreover, we can zoom the low-current part in Fig.~\ref{fig:Simple_Case_Region_Overall} to be Fig.~\ref{fig:Simple_Case_Region_Contour}. Through drawing contouring lines, we can find the hosting capacity region can shrink into a convex one with current level threshold decreasing. Regarding a practical 402A threshold, it is low enough to ensure a convex respective hosting capacity region.

\begin{figure}
    \centering
    \includegraphics[width=0.7\linewidth]{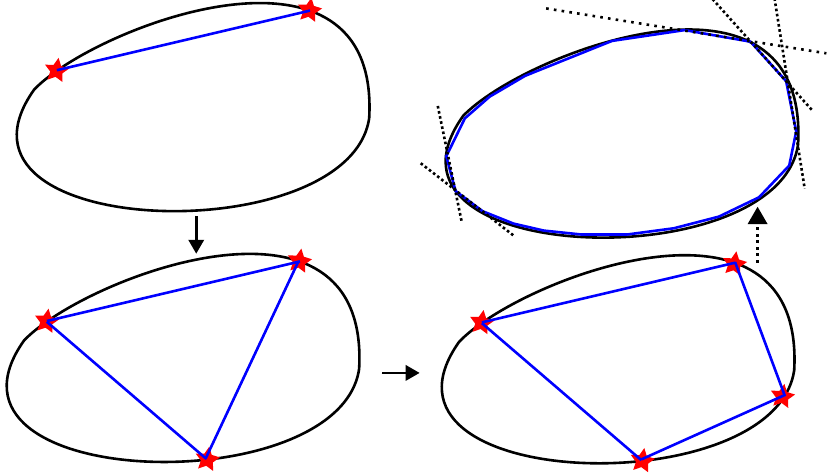}
    \caption{Convex region linearization}
    \label{fig:Simple_Case_Convex}
\end{figure}
Now, you may be confused why we so desire the convexity. To answer it, we will give some brief explanations here, and more details will be provided later in coming sections. As shown in Fig.~\ref{fig:Simple_Case_Convex}, the region benefits from convexity due to the feasibility guarantee for the connection line between two existing feasible points. Based on three feasible points, we can derive a feasible planar triangle region accordingly. With emerging known feasible points, the derived feasible region comes more and more close to the original one, and the derived one can be easily identified by a limited number of linear constraints, which is more friendly for numerical processing. That explains why we so desire the convexity.  

In summary, the hosting capacity region defined by DistFlow cannot be ensured convex, while such convexity can still be met in some specific scenarios. The convexity can benefit our future hosting capacity region exploitation, and we should take it seriously.

\subsubsection{Testing Case Explanation}
\label{subsubsec:testing_case}
\begin{figure}
    \centering
    \includegraphics[width=0.7\linewidth]{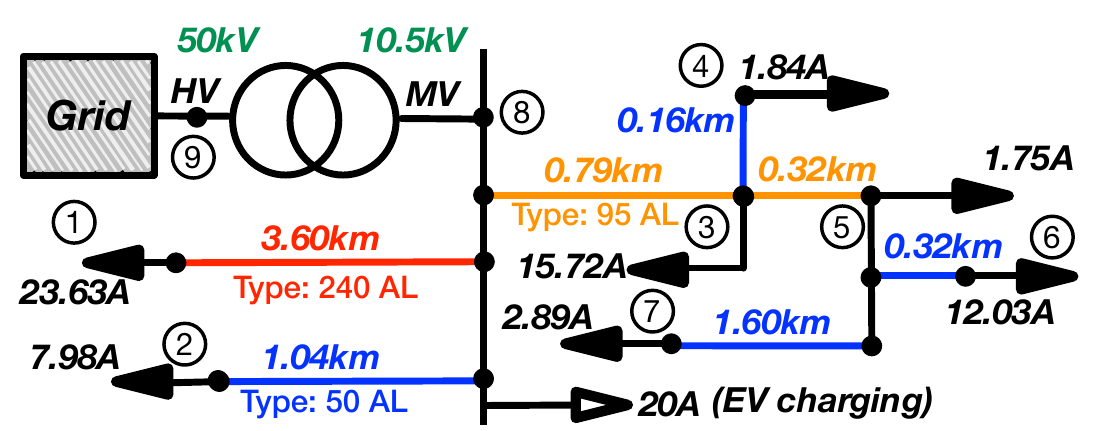}
    \caption{Testing Case Schematic}
    \label{fig:testing_case_schematic}
\end{figure}

Unfortunately, a small grid case in the previous section can be hard to convince DSOs, as a practical grid case usually owns more buses. In this chapter, we will use a branch of standard Dutch MV grid case as a common testing case in coming sections. The grid schematic has been given in Fig.~\ref{fig:testing_case_schematic}, where original loads share the same power factor of 0.98. Bus~9 is assumed a slack bus.

The EV charging station is connected to Bus 8 and its nominal charging current is set 20A. However, considering bidirectional charging converter technology and "Vehicle to Grid" policy, through taking Bus 8 as a specific POC, we will try to exploit its hosting capacity region. Bus~7 is selected as well to assess their combinational feasible region. Such region will provide an important reference for converter capacity redesigning and policy adjustment. Cable and transformer impedance are listed in Table~\ref{tab:grid parameter}.

\begin{table}[ht]
    \centering
    \caption{Power cable and transformer parameters}
    \begin{tabular}{c c c}
        \toprule
        Object  & Resistance & Reactance (50Hz)  \\
        \midrule
          240 AL Cable&  126m$\Omega$/km & 116m$\Omega$/km\\
          95 AL Cable &  320m$\Omega$/km & 188m$\Omega$km\\
          50 AL Cable&  641m$\Omega$/km & 204m$\Omega$/km\\
         36MVA Transformer (50kV/10.5kV) &  0.0022p.u.  & 0.065p.u. \\
         \bottomrule
    \end{tabular}
    \label{tab:grid parameter}
\end{table}

\subsection{Linearized DistFlow model}
\subsubsection{Model Explanation}
Our intuitive to deal with such imperfect reality is rational simplification. In accordance with \cite{baran1989network}, a linearized DistFlow model has been proposed as shown in (\ref{eq:linear_model}). Compared to (\ref{eq:tight_model}), we use some cancelling symbols to represent how such linearized model is derived. Especially, since $l_{ij}$-related items have been ignored in (\ref{eq:linear_model_1}-\ref{eq:linear_model_3}), (\ref{eq:linear_model_4}) can be rephrased as an inequality constraint directly. (\ref{eq:linear_model_4}) is a nonlinear while convex constraint. The reason why we still name such model "Linearized DistFlow Model" is that all equality constraints become linear now.

Since all constraints are convex, regarding (\ref{eq:linear_model}), its respective feasible region is theoretically convex. Such convexity can powerfully support our exploitation on hosting capacity region as stated above. Moreover, such linearization makes senses when $l_{ij}$ is comparatively low, and as shown in Fig.~\ref{fig:Simple_Case_Region_Contour}, the region is naturally convex with low $l_{ij}$ in a simple grid case.

\begin{subequations}
\begin{equation}
  \label{eq:linear_model_1}
    P_{ij}\simeq\sum_{k: (j,k)\in E } P_{jk}\cancel{+r_{ij}l_{ij}}-P_j
\end{equation}    
\begin{equation}
  \label{eq:linear_model_2}
    Q_{ij}\simeq\sum_{k: (j,k)\in E } Q_{jk}\cancel{+x_{ij}l_{ij}}-Q_j 
\end{equation}
\begin{equation}
  \label{eq:linear_model_3}
    v_j \simeq v_i-2(r_{ij}P_{ij}+x_{ij}Q_{ij})\cancel{+(r_{ij}^2+x_{ij}^2)l_{ij}}
\end{equation}
\begin{equation}
  \label{eq:linear_model_4}
    \cancel{l_{ij} \leq l_{ij}^{max}},~ \cancel{l_{ij} =} \frac{P_{ij}^2+Q_{ij}^2}{v_i}\leq l_{ij}^{max}
\end{equation}
\begin{equation}
  \label{eq:linear_model_5}
    P_i^{min}\leq P_i \leq P_i^{max}
\end{equation}
\begin{equation}
  \label{eq:linear_model_6}
    Q_i^{min}\leq Q_i \leq Q_i^{max}
\end{equation}
\begin{equation}
  \label{eq:linear_model_7}
    v_{i}^{min}\leq v_i \leq v_{i}^{max}
\end{equation}
\label{eq:linear_model}
\end{subequations}

\subsubsection{Heuristic Convex Hull Algorithm}
The basic idea of convex region exploitation has been explained above. Meanwhile, it still has not explained how we can define those boundary points efficiently. There are several classic convex hull determination algorithms, and many researchers have investigated a lot in this topic \cite{avis1997good}. However, it seems irrational to employ these algorithms directly, as they are designed for a group of known points. In our case, instead of randomly choosing testing points, we can manage how testing points are generated, and we can use such capability to realize heuristic region assessment.

\begin{figure}
    \centering
    \includegraphics[width=0.9\linewidth]{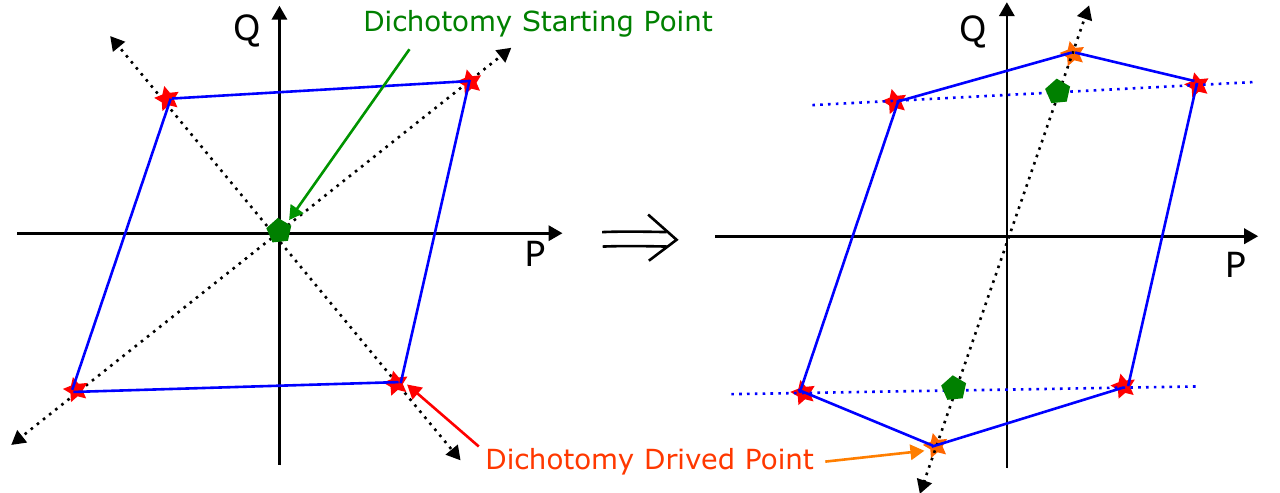}
    \caption{Hosting capacity region exploitation flow chart }
    \label{fig:Linear_DistFlow_FlowChart_2D}
\end{figure}

Regarding the scalar value boundary exploitation, Dichotomy method in {Algorithm~\ref{Alg:Dichotomy_SinglePOC_HC}} is intuitive and efficient. Based on such algorithm, we can continue from a simple case, where the hosting capacity region is limited in 2D. The corresponding heuristic exploitation flow chart has been given in Fig.~\ref{fig:Linear_DistFlow_FlowChart_2D}. Through selecting two initial linear trajectories, we can determine four boundary points by employing Algorithm~\ref{Alg:Dichotomy_SinglePOC_HC}. 
In the initial stage, all lower-bound starting points are set zero point, as an initial status without any extra energy unit integration is assumed acceptable. Later on, we will select another direction, while the lower-bound starting point in Algorithm~\ref{Alg:Dichotomy_SinglePOC_HC}  can be updated as the conjunction point between the connection line and this new linear trajectory. By iteratively running Algorithm~\ref{Alg:Dichotomy_SinglePOC_HC}, the region will expand continuously. Through converting these connection lines as linear constraints, the hosting capacity region will be algebraically represented. 

\begin{algorithm}[H]
\KwData{Lower-bound starting point $S_l$, upper-bound starting point $S_u$}
\KwResult{Derived boundary point $S$}
$S_a \gets S_l$\ , $S_b \gets S_u$\; %\Comment*[r]{Initialize bounds}
\text{Set stopping criteria} $\epsilon$ \;
\Repeat{$|S_b-S_a| \leq \epsilon$}
{ $S \gets \frac{S_a + S_b}{2}$ \;
  Run grid simulation programs for testing\;
  \uIf{$S$ \text{is acceptable}}{ 
  $S_a \gets \frac{S_a + S_b}{2}$ \; } %\Comment*[r]{Voltage and current constraints are met}}
    %\uElseIf{}{}
    \Else{
    $S_b \gets \frac{S_a + S_b}{2}$ \;}
    }
\caption{Dichotomy Algorithm}
\label{Alg:Dichotomy_SinglePOC_HC}
\end{algorithm}

\begin{algorithm}
\caption{Linearized Hosting Capacity Region Exploitation in $\textbf{R}^{2}$}
\label{Alg:High_Dimension_Capacity_Linear_2D}
\KwData{Orthogonal vector basis set $\textbf{V}=\{\textbf{v}_1,\textbf{v}_2\} $}
%( Preliminary~2: $\textbf{v}_i \textbf{v}_j^T=\textbf{v}_j \textbf{v}_k^T= \textbf{v}_k \textbf{v}_i^T$ for any $\textbf{v}_i, \textbf{v}_j, \textbf{v}_k \in \textbf{V}$ )\\
\KwResult{Half space set $\textbf{H}$}
\text{Set stopping criteria} $\epsilon$ \;
Run \textbf{Dichotomy} to find bounds  $\alpha_{1+}\cdot \textbf{v}_1$ and  $\alpha_{1-} \cdot \textbf{v}_1$ \;
$\textbf{v}_{1+} \gets \alpha_{1+}\cdot \textbf{v}_1$ , $\textbf{v}_{1-} \gets \alpha_{1-}\cdot \textbf{v}_1$ \;
Run \textbf{Dichotomy} to find bounds  $\alpha_{2+}\cdot \textbf{v}_2$ and  $\alpha_{2-} \cdot \textbf{v}_2$ \;
$\textbf{v}_{2+} \gets \alpha_{2+}\cdot \textbf{v}_2$ , $\textbf{v}_{2-} \gets \alpha_{2-}\cdot \textbf{v}_1$ \;
Initialize Sequence $\textbf{Q} \gets [~\textbf{v}_{1+},~\textbf{v}_{2+}~,\textbf{v}_{1-},~\textbf{v}_{2-}~]$\;
\Repeat{ size($\textbf{Q}$) $\geq \epsilon$}{
Initialize integer number $i$ < size($\textbf{Q}$)\;
Select the $i$th and $(i+1)$th element in $\textbf{Q}$, denoted by $\textbf{v}_i$, $\textbf{v}_{i+1}$ \;
( if $i+1 >$ size($\textbf{Q}$), $\textbf{v}_{i+1} \gets \textbf{v}_{1}$ )\\
Initialize $\textbf{v}_0$ as a average of $\textbf{v}_i$ and $\textbf{v}_{i+1}$ \;
Run \textbf{Dichotomy} to find bounds  $\alpha_{0+}\cdot \textbf{v}_0$ and  $\alpha_{0-} \cdot \textbf{v}_0$ \;
$\textbf{v}_{0+} \gets \alpha_{0+}\cdot \textbf{v}_0$ , $\textbf{v}_{0-} \gets \alpha_{0-}\cdot \textbf{v}_0$ \;
\lIf{i > size($\textbf{Q}$)/2}
{j$\gets$ i-size($\textbf{Q}$)/2+1}
\lElse{j$\gets$ i+size($\textbf{Q}$)/2+1}  
Insert $\textbf{v}_{0+}$ between $\textbf{v}_i$ and $\textbf{v}_{i+1}$ in $\textbf{Q}$ \;
Insert $\textbf{v}_{0-}$ between $\textbf{v}_j$ and $\textbf{v}_{j+1}$ in $\textbf{Q}$ \;
}
Initialize half space set $\textbf{H} \gets \emptyset$\;
\For{$\textbf{v}_i$ in $\textbf{Q}$}{
    Use $\textbf{v}_i$ and $\textbf{v}_{i+1}$ to generate half space $h_{i}$ \;
    ( if $i+1 >$ size($\textbf{Q}$), $\textbf{v}_{i+1} \gets \textbf{v}_{1}$ )\\
    Add $h_{i}$ to $\textbf{H}$\;
}
\end{algorithm}

Meanwhile, the convex hull should be represented by those boundary lines. When a boundary pair in one direction is derived, it still needs to figure out how to connect these lines. How we can heuristically calculate cutting lines determines our assessment efficiency. As shown in Fig.~\ref{fig:Linear_DistFlow_FlowChart_2D}, with two more boundary points (orange) generated, top and bottom lines in the left quadrangle should be replaced by four other lines in the right hexagon. So we can use a list to denote the connection sequence of all points. When a new pair is derived, we can insert them appropriately to the previous list. Inspired by such principle,  Algorithm~\ref{Alg:High_Dimension_Capacity_Linear_2D} is written down to illustrate the whole procedure of 2D hosting capacity region assessment. 

\subsubsection{Region Correction}
Since the linearized DistFlow model acquires convexity with the sacrifice of its accuracy, even through Algorithm~\ref{Alg:High_Dimension_Capacity_Linear_2D}, power engineers are still hesitant to totally trust the results. Commonly, the grid operation reliability is the priority, and the DSO prefers to see that the derived region from the linearized model keeps feasible even with region space reduction. In the other word, we should avoid letting the linearized and accurate region overlap, and the linearized one must keep totally covered by the accurate one.

From the perspective of active power, (\ref{eq:linear_model_1}) is recycled and we repeat it as (\ref{eq:linear_model_1_recite}) for better reading experience. Compared to the accurate model, the linearized one provides an under-estimator of injection power $P_j$.  
\begin{equation}
    \label{eq:linear_model_1_recite}
    P_{ij}\simeq\sum_{k: (j,k)\in E } P_{jk}\cancel{+r_{ij}l_{ij}}-P_j
\end{equation}

Through using $\Hat{P_j}$ to denote the linearized one, it can be written as 
\begin{equation}
    \Hat{P_j} = P_j -r_{ij}l_{ij} \leq P_j 
\end{equation}
Although $l_{ij}$ is dynamic in various scenarios, considering $l_{ij}$ is natrually constrained, we can still derive
\begin{equation}
\label{eq:Linearized_Region_Correction}
  \Hat{P_j} \leq P_j \leq \Hat{P_j}+r_{ij}l_{ij}^{max}  
\end{equation}
Therefore, if $\mathbcal{C}$ and $\Hat{\mathbcal{C}^{}}$ are used to denote the exact and linearized hosting capacity region, we can conclude that
\begin{equation}
  \bigcap_{(i,j)\in E } \{~\mathbcal{C} \medcap (\mathbcal{C}+r_{ij}l_{ij}^{max}  ) \medcap (\mathbcal{C}+x_{ij}l_{ij}^{max}  )~\} \subseteq \mathbcal{R}
  \label{eq:Linearized_Region_Intersection}
\end{equation}
As soon as $P_j$ is selected as a dimension of $\mathbcal{C}$, $\mathbcal{C}+r_{ij}l_{ij}^{max}$ denotes a new region generated by shifting $\mathbcal{C}$ along the axis of $P_j$ by $r_{ij}l_{ij}^{max}$, otherwise it can be taken as a copy of $\mathbcal{C}$ directly. $\mathbcal{C}+x_{ij}l_{ij}^{max}$ shares a similar definition. 

Therefore, we can employ region correction following the principle of (\ref{eq:Linearized_Region_Intersection}). The corrected region is ensured totally covered by the exact one, eventually becoming acceptable for power engineers as stated above.

\subsubsection{Case Study}
\label{subsubsec:case_2D_linear}
\begin{figure}
     \centering
     \begin{subfigure}[b]{0.49\textwidth}
         \centering
         \includegraphics[width=\textwidth]{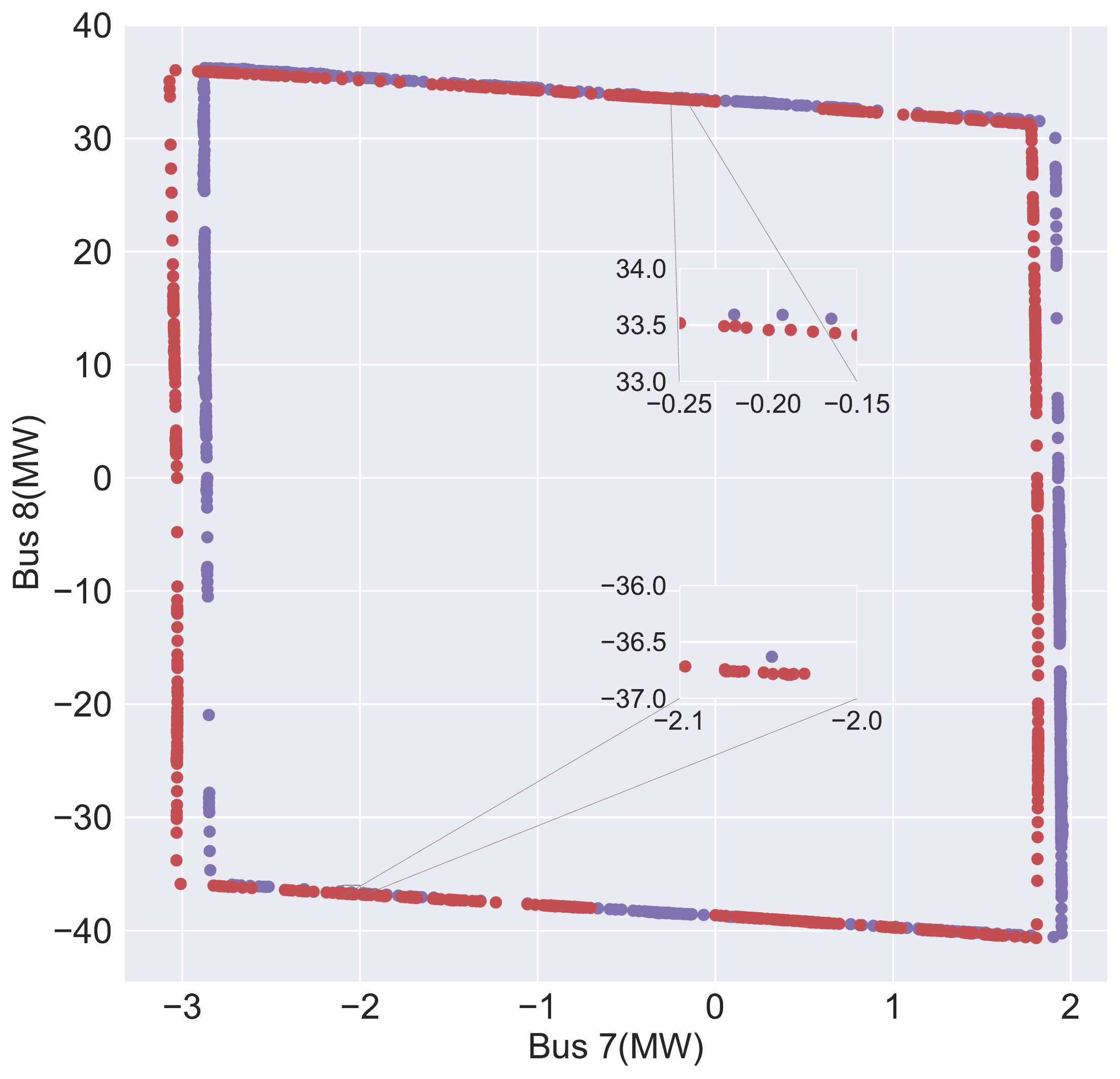}
         \caption{Original hosting capacity region}
         \label{fig:Linear_DistFlow_Region_Compare_Bad}
     \end{subfigure}
     \hfill
     \begin{subfigure}[b]{0.49\textwidth}
         \centering
         \includegraphics[width=\textwidth]{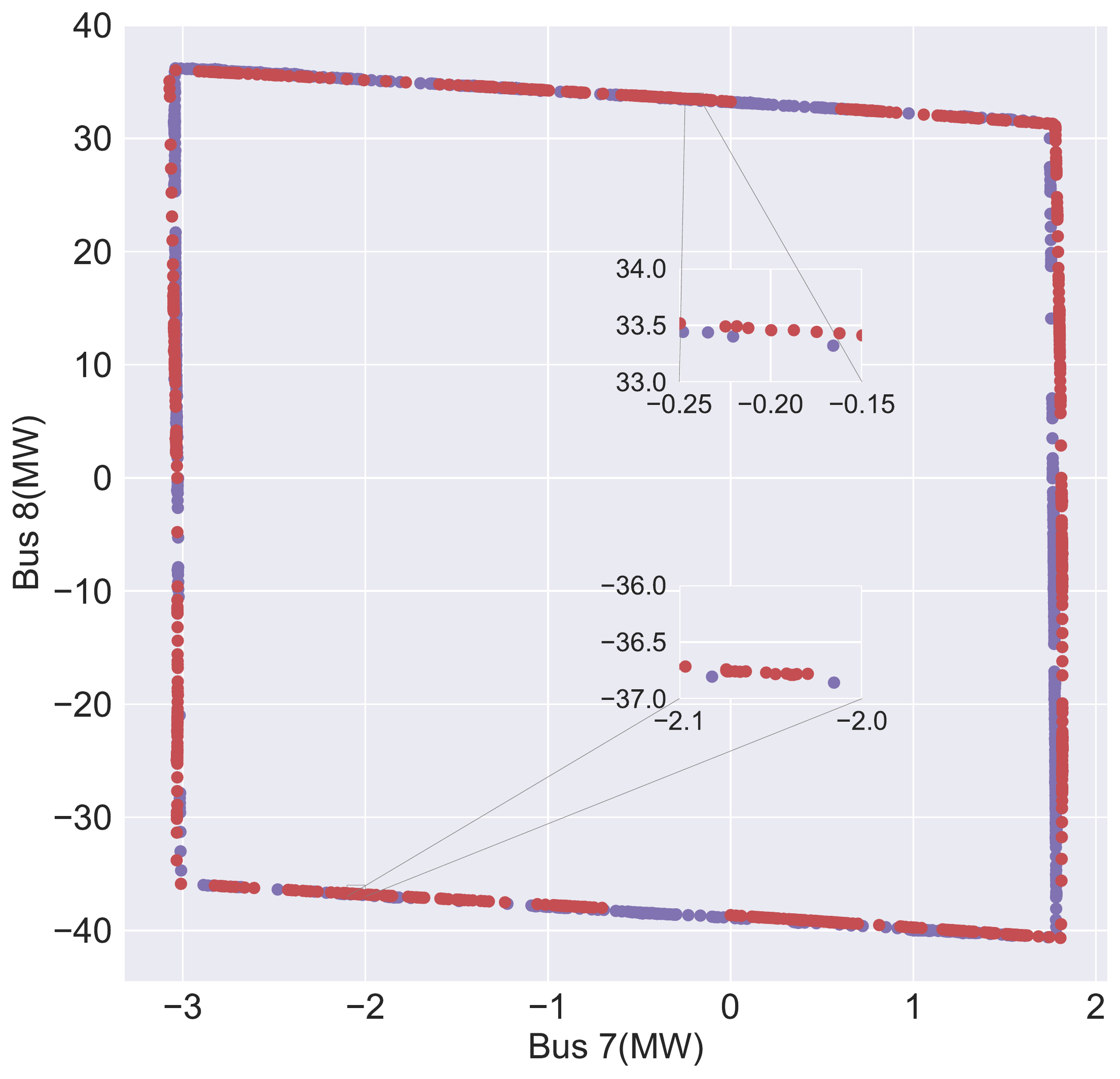}
         \caption{Shifted hosting capacity region}
         \label{fig:Linear_DistFlow_Region_Compare_Good}
     \end{subfigure}
        \caption{Hosting capacity region boundary points}
\end{figure}

Using the testing case in Fig.~\ref{fig:testing_case_schematic} at Section~\ref{subsubsec:testing_case}, with only active power inspected, we can derive the hosting capacity region regarding Bus~7 and Bus~8. Through running Algorithm~\ref{Alg:High_Dimension_Capacity_Linear_2D}, Fig.~\ref{fig:Linear_DistFlow_Region_Compare_Bad} is derived, where purple ones are boundary points derived from linearized model, and red ones are from exact model. Without region correction, these two regions overlap in both axis directions. Referring to (\ref{eq:Linearized_Region_Intersection}), the hosting capacity region (purple) from linearized model has been shifted as shown in Fig.~\ref{fig:Linear_DistFlow_Region_Compare_Good}. Through zooming out vertical axis, it still can be seen that the shifted (purple) and exact (red) region overlap. 

\begin{figure}
     \centering
     \begin{subfigure}[b]{0.47\textwidth}
         \centering
         \includegraphics[width=\textwidth]{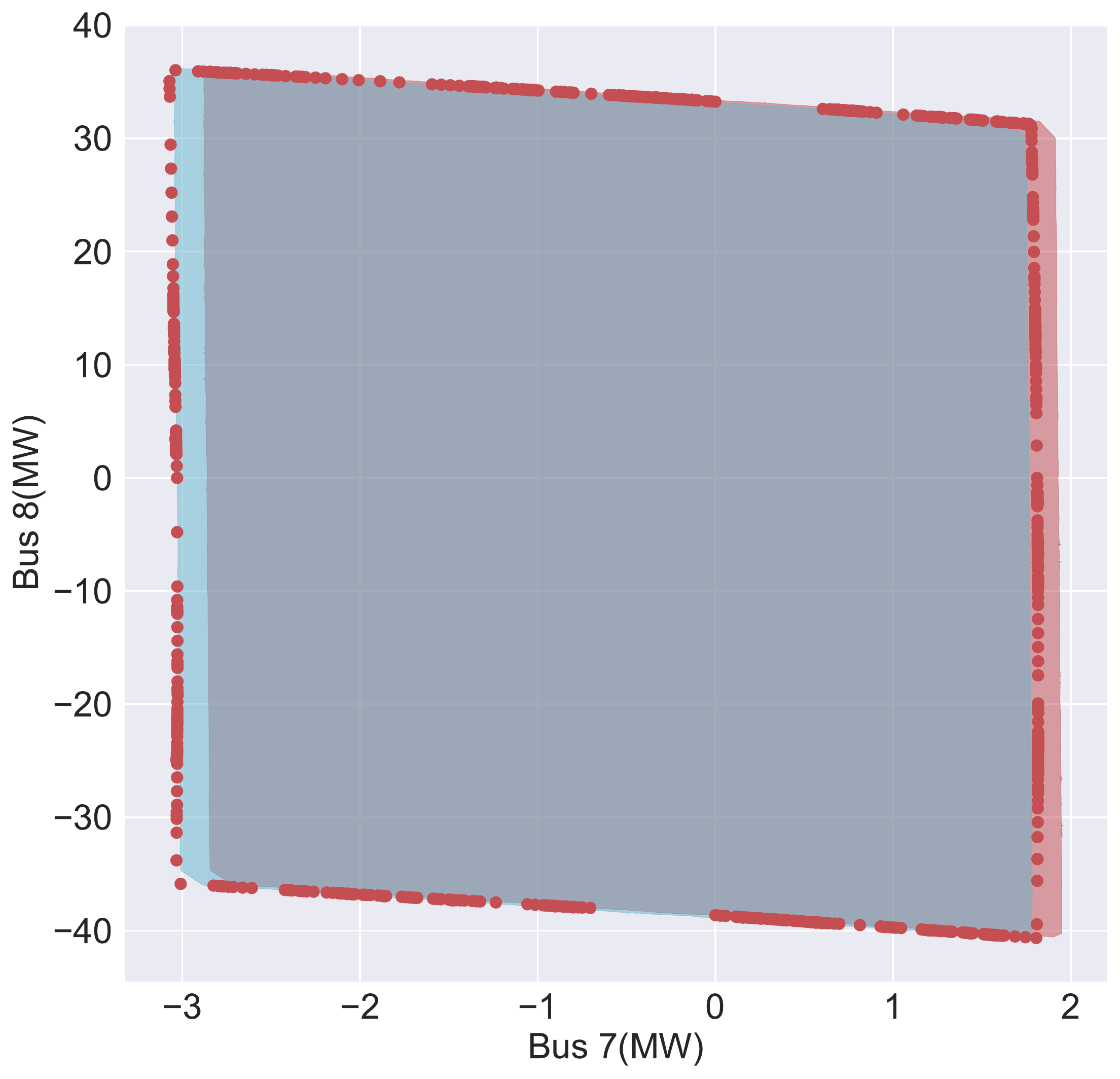}
         \caption{Updated hosting capacity region}
         \label{fig:Linear_DistFlow_Region_Compare_Correct}
     \end{subfigure}
     \hfill
     \begin{subfigure}[b]{0.49\textwidth}
         \centering
         \includegraphics[width=\textwidth]{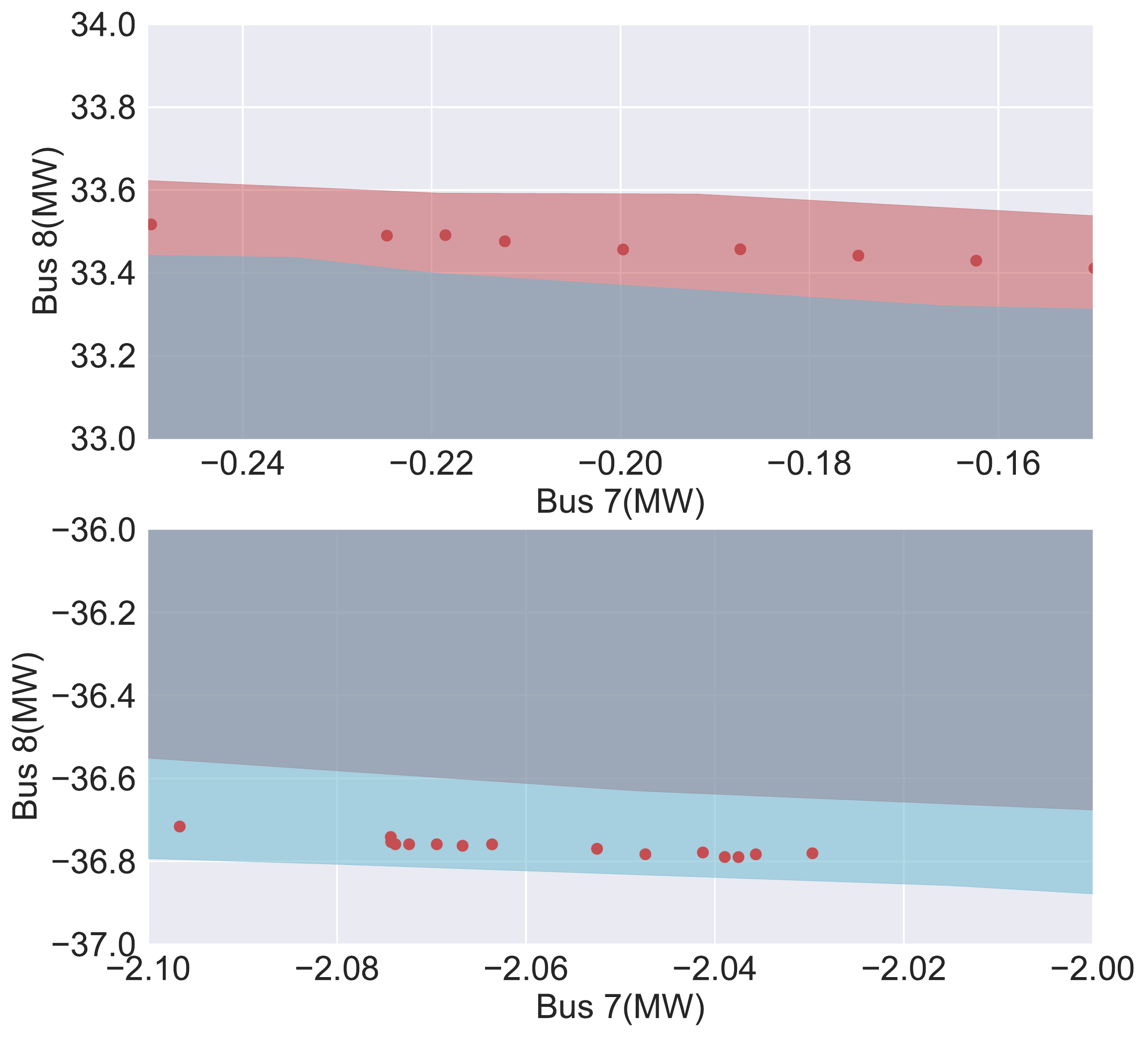}
         \caption{Zoomed hosting capacity region}
         \label{fig:Linear_DistFlow_Region_Compare_Zoom}
     \end{subfigure}
        \label{fig:Linear_DistFlow_Region_Compare_Correct_Zoom}
        \caption{Hosting capacity region correction}
\end{figure}

As shown in Fig.~\ref{fig:Linear_DistFlow_Region_Compare_Correct}, the light blue and light red polygons are original and shifted hosting capacity region. The grey part is the intersection of both regions, which is adopted as the updated hosting capacity region referring to (\ref{eq:Linearized_Region_Intersection}). The good news is that this updated one is almost totally covered by the exact one (red points). Moreover, through zooming the boundary part, we can still figure out such covering relationship. Eventually, the final hosting capacity region denoted by grey part is assumed convincing and reliable to power engineers, and the proposed hosting capacity region assessment technology is confirmed valid.

\subsection{Relaxed DistFlow model}
\subsubsection{Model Description}
As explained above, the linearized DistFlow model can contribute to a valid hosting capacity region, while with potential region space sacrifice. Meanwhile, referring to \cite{low2014convex}, there seems another route to utilize model convexity for hosting capacity region assessment. In advance of further discussions, for convenience of model description, we can denote an exact solution by $\mathbcal{S}$ and the exact feasible region by $\mathbcal{R}$ constrained by (\ref{eq:tight_model}) as given in (\ref{eq:solultion_define}). Thus we can provide an alternative expression about hosting capacity region $\mathbcal{C}$ as illustrated in (\ref{eq:hostcap_define}), which is derived from hosting power values in $\mathbcal{R}$. Each element in $\mathbcal{C}$ responds to a specific $\mathbcal{S}$, while only bus injection power is concerned.
\begin{equation}
\label{eq:solultion_define}
\begin{aligned}
    \mathbcal{S} &\equiv \{P_{ij},Q_{ij},P_j,Q_j,v_{j},l_{ij}~|~(i,j)\in E,~ \text{(1) is met}\} \\
    \mathbcal{R} &\equiv \{\mathbcal{S}~|~\mathbcal{S}~\text{meets~\text{DistFlow Model}} \}
\end{aligned}
\end{equation}
\begin{equation}
\label{eq:hostcap_define}
    \mathbcal{C} \equiv\textbf{\{}~\{P_j,Q_j~|~(i,j)\in E\}~\boldsymbol{|}~\mathbcal{R}\neq \emptyset~\textbf{\}}
\end{equation}

Convex relaxation on DistFlow model has been deeply investigated in the topic of optimal power flow calculation, which has been written in (\ref{eq:relaxed_model}). Instead, in this chapter, we focus on its corresponding feasible region. After relaxation, the new feasible region $\mathbcal{R}^*$ becomes convex where $\mathbcal{R} \subseteq \mathbcal{R}^*$. The relaxation is highlighted by red in (\ref{eq:tight_model_4}).
\begin{subequations}
\begin{align}
    P_{ij}&=\sum_{k: (j,k)\in E } P_{jk}+r_{ij}l_{ij}-P_j \\
    Q_{ij}&=\sum_{k: (j,k)\in E } Q_{jk}+x_{ij}l_{ij}-Q_j \\
    v_j &=v_i-2(r_{ij}P_{ij}+x_{ij}Q_{ij})+(r_{ij}^2+x_{ij}^2)l_{ij}\\
    l_{ij}& {\geq} \frac{P_{ij}^2+Q_{ij}^2}{v_i}\\
    P_i^{min}&\leq P_i \leq P_i^{max}\\
    Q_i^{min}&\leq Q_i \leq Q_i^{max}\\
    v_{i}^{min}&\leq v_i \leq v_{i}^{max}\\
    &~~~~ l_{ij} \leq l_{ij}^{max}
\end{align}
\label{eq:relaxed_model}
\end{subequations}
Most new generated points will  meet all constraints in (\ref{eq:tight_model}) except (\ref{eq:tight_model_4}), which means $\mathbcal{S}^* \not\in \mathbcal{R}$. Equation~(\ref{eq:violation}) is given for further illustration, where the subscript $^*$ to distinguish elements in $\mathbcal{S}^*$.
\begin{equation}
    l^*_{ij}>\frac{{P^*_{ij}}^2+{Q^*_{ij}}^2}{v^*_i}
    \label{eq:violation}
\end{equation}

As mentioned before, $\mathbcal{S}^*$ is still infeasible, thus we would like to map a certain $\mathbcal{S}^*$ to another feasible solution $\Tilde{\mathbcal{S}^{}}$, indicating $\Tilde{\mathbcal{S}^{}} \in \mathbcal{R}$. Through shifting $\mathbcal{S}^*$, we can regenerate another point defined as (\ref{eq:new_point}), where the hat symbol distinguishes elements of the regenerated solution $\Tilde{\mathbcal{S}^{}}$. $\epsilon$ is an independent scalar variable. With a suitable value for $\epsilon$, $\Tilde{\mathbcal{S}^{}}$ can be confirmed to meet all constraints in (\ref{eq:tight_model}), indicating a new exact feasible solution while without repetitive powerflow calculation.

\begin{subequations}
\begin{align}
    \Tilde{v}_i= v^*_i,\Tilde{v}_j= v^*_j \\
    \Tilde{l}_{ij}= {l}^*_{ij}-\epsilon \\
    \Tilde{P}_{ij}= {P}^*_{ij}-r_{ij}\epsilon/2 \\
    \Tilde{Q}_{ij}= {Q}^*_{ij}-x_{ij}\epsilon/2 \\
    \Tilde{P}_{i}= {P}^*_{i}-r_{ij}\epsilon/2 \\
    \Tilde{P}_{j}= {P}^*_{j}-r_{ij}\epsilon/2 \\
    \Tilde{Q}_{i}= {Q}^*_{i}-x_{ij}\epsilon/2 \\
    \Tilde{Q}_{j}= {Q}^*_{j}-x_{ij}\epsilon/2
\end{align}
\label{eq:new_point}
\end{subequations}

\subsubsection{Point-wise Region Assessment}
The proper value determination of $\epsilon$ is given in Algorithm~\ref{Ag:epsilon_value}, where Gauss-Seidel method is adopted to ensure a stable solution. The bus number of such EV charging infrastructure system is denoted by $N$, then Algorithm~\ref{Ag:epsilon_value} will run $N-1$ times due to $N-1$ links. In a distributive computation scheme, at most $N-1$ processor can run Algorithm~\ref{Ag:epsilon_value} simultaneously as each $\epsilon$ will respond to a certain link independently. 

\begin{algorithm}[!h]
\SetAlgoLined
\KwResult{ $\epsilon$}
Initialize $\Tilde{v}_i= v^*_i,~\Tilde{P}_{ij}^{(0)}= {P}^*_{ij},~\Tilde{Q}_{ij}^{(0)}= {Q}^*_{ij},~\Tilde{l}_{ij}^{(0)}= {l}^*_{ij}$\;
Initialize $\epsilon^{(0)}= \Tilde{l}_{ij}^{(0)}-[(\Tilde{P}_{ij}^{(0)})^2+(\Tilde{Q}_{ij}^{(0)})^2]/\Tilde{v}_i$, $n=0$\;
\While{$|\epsilon^{(i)}|$ is larger than threshold}{
$n\leftarrow n+1$,
$\Tilde{l}_{ij}^{(n)}\leftarrow \Tilde{l}_{ij}^{(n-1)}-\epsilon^{(n-1)}$\;
$\Tilde{P}_{ij}^{(n)}\leftarrow \Tilde{P}_{ij}^{(n-1)}-r_{ij}\epsilon^{(n-1)}/2$\;
$\Tilde{Q}_{ij}^{(n)}\leftarrow \Tilde{Q}_{ij}^{(n-1)}-x_{ij}\epsilon^{(n-1)}/2$\;
$\epsilon^{(n)}\leftarrow  \Tilde{l}_{ij}^{(n)}-[(\Tilde{P}_{ij}^{(n)})^2+(\Tilde{Q}_{ij}^{(n)})^2]/\Tilde{v}_i$\;
}
$\epsilon \leftarrow {l}^*_{ij}-\Tilde{l}_{ij}^{(n)} $\;
\caption{$\epsilon$ value determination}
\label{Ag:epsilon_value}
\end{algorithm}

\begin{figure}
    \centering
    \includegraphics[width=\linewidth]{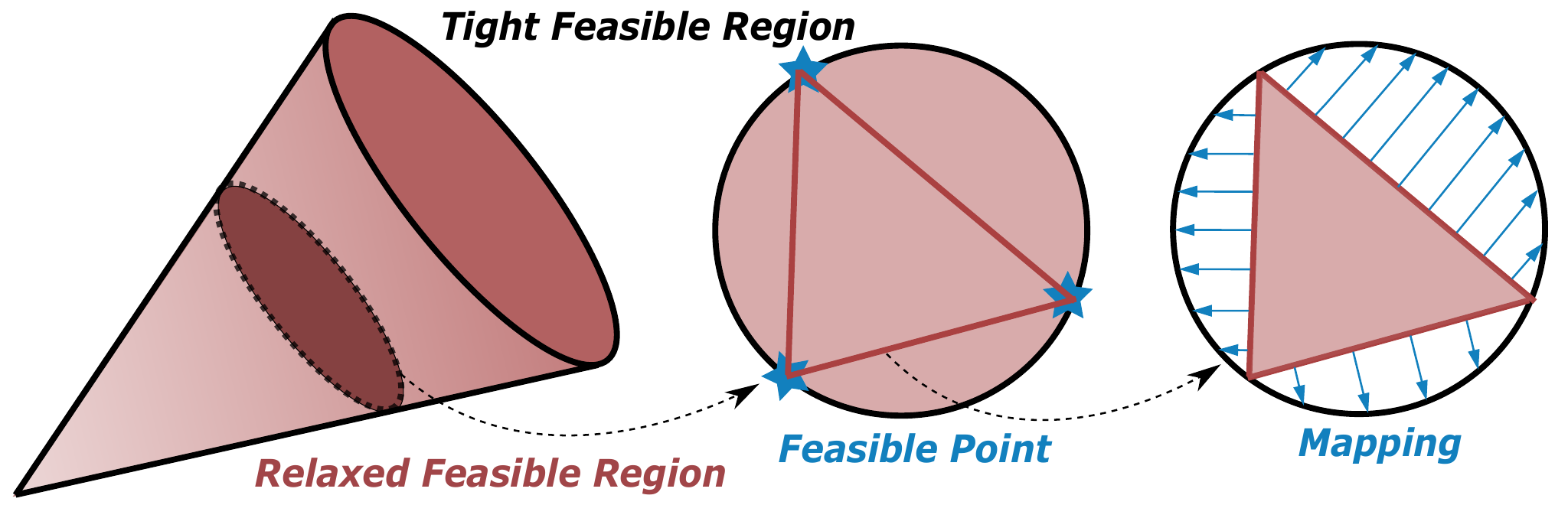}
    \caption{Hosting capacity region exploitation principle }
    \label{fig:principle}
\end{figure}

The principle of capacity reconfiguration is illustrated in Fig.~\ref{fig:principle}. Equation (4d) is a second-order conic constraint, whose respective feasible region is a cone as represented in Fig.~\ref{fig:principle}. The surface of this cone represents the exact feasible region defined by tight constraint (\ref{eq:tight_model_4}). If we look into its crossing section, all feasible solutions $\mathbcal{S}$ calculated by powerflow calculation tools are only possible to be located upon the bound of this circle, which is denoted by the circular bold black line. Through connecting these feasible points, we can derive several lines. Except endpoints, these lines are out of the tight feasible region while still belong to the relaxed one. In accordance with (\ref{eq:new_point}), we will calculate $\epsilon$ and map infeasible points to the circular bold line, eventually deriving a new set of feasible points and realizing hosting capacity region exploitation.

Now, you may also notice that such method is point-wise, indicating we cannot directly figure out extra linear boundaries to help formulate a polygon, which is essential to represent the hosting capacity region. Instead, we are using its convexity to generate a point very close to its respective true solution. Moreover, with distributive computation capability of such method, repetitive  power flow calculation can be avoided, replaced by parallel $\epsilon$ computation in (\ref{eq:new_point}) using Algorithm~\ref{Ag:epsilon_value}. Therefore, after running Algorithm~\ref{Alg:High_Dimension_Capacity_Linear_2D} with several boundary points derived, we can take it as a fast-computing interpolation method when we want to derive more boundary points to represent a more precise hosting capacity feasible region.

\subsubsection{Case Study}
\label{subsubsec:case_interpolation}
Still focusing on the testing case and results in Section.~\ref{subsubsec:case_2D_linear}, we can simplify the hosting capacity region as a quadrilateral defined by four vertices as listed in Table~\ref{tab:region_vertex}. Each point is on the edge of grid constraint violation, indicated by relatively low or high Bus~7 voltage, especially in the context of full transformer loading. Therefore, combining with previous geometrical results in Fig.~\ref{fig:Linear_DistFlow_Region_Compare_Correct}, we select these four vertices to realize interpolation as illustrated above. 

However, only four vertices derived from powerflow calculation may not be sufficient to get the whole region due to natural accompanying reactive power correction in Algorithm~\ref{Ag:epsilon_value}.When $\epsilon$ is derived, it imposes changes on $P_i$ and $Q_i$ simultaneously. However, in this case, We aims for a combinational feasible region only related to active power, and extra reactive power load is expected to keep zero. If the reactive power correction is huge, even we derive another feasible point successfully, it cannot be assumed an acceptable interpolated point.
\begin{table}
    \centering
    \begin{tabular}{c c c c}
    \toprule
       No.  &  Bus~7/8 Load~(MW) & Bus~7 Voltage (p.u.) & Transformer Load (\%) \\
       \midrule
        1 & ~1.594 / 31.484 & 0.901 & 100.064\\
        2 & -3.039 / 36.026  & 1.099 & 99.997\\
        3 & -3.032 / -24.963 & 1.099 & 100.009\\
        4 & ~1.805 / -40.567 & 0.900 & 99.742\\
        \bottomrule
    \end{tabular}
    \caption{Table of vertices in hosting capacity region}
    \label{tab:region_vertex}
\end{table}

\begin{figure}
     \centering
     \begin{subfigure}[b]{0.47\textwidth}
         \centering
         \includegraphics[width=\textwidth]{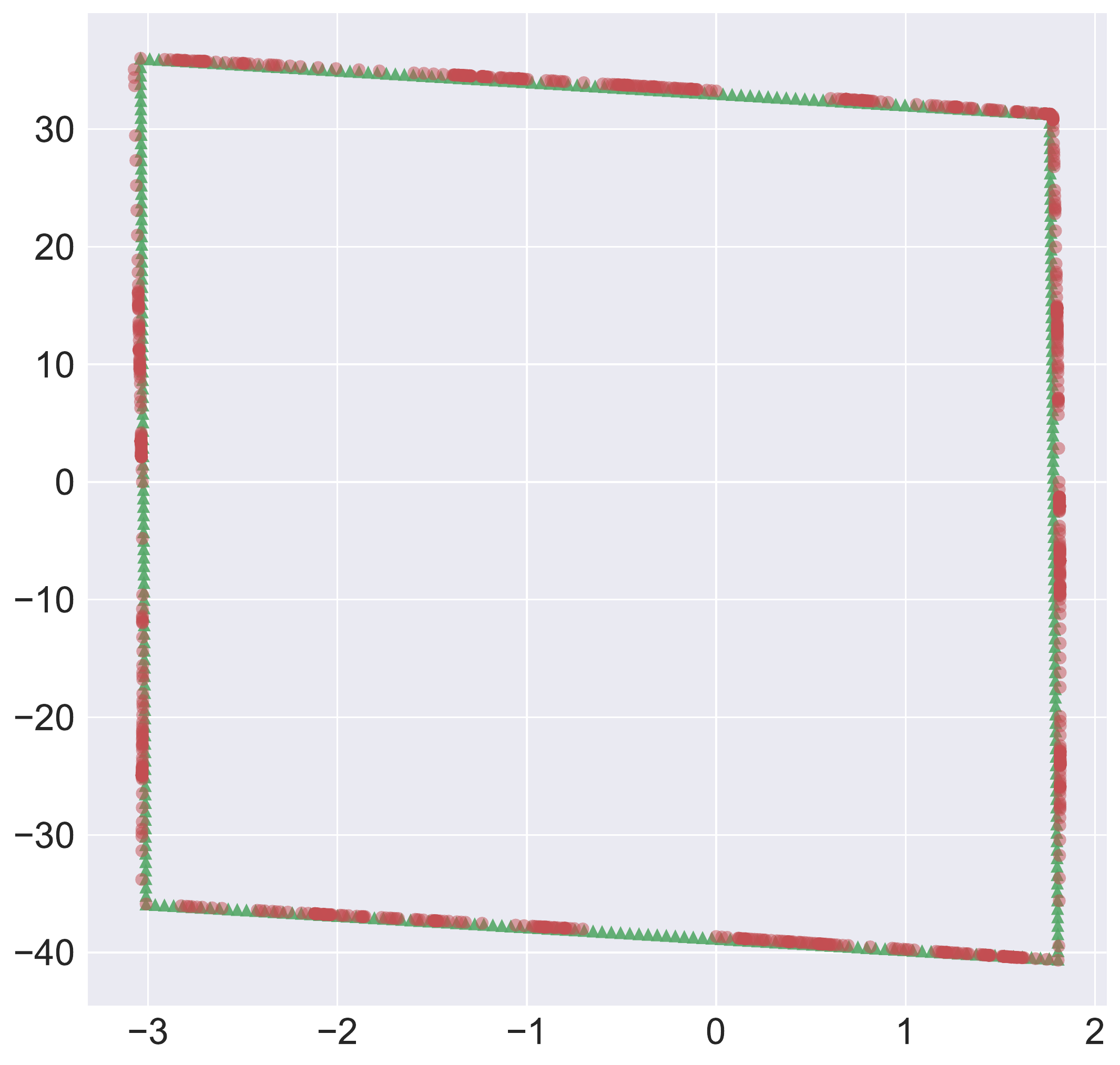}
         \caption{Updated hosting capacity region}
         \label{fig:Interpolation_Overall}
     \end{subfigure}
     \hfill
     \begin{subfigure}[b]{0.49\textwidth}
         \centering
         \includegraphics[width=\textwidth]{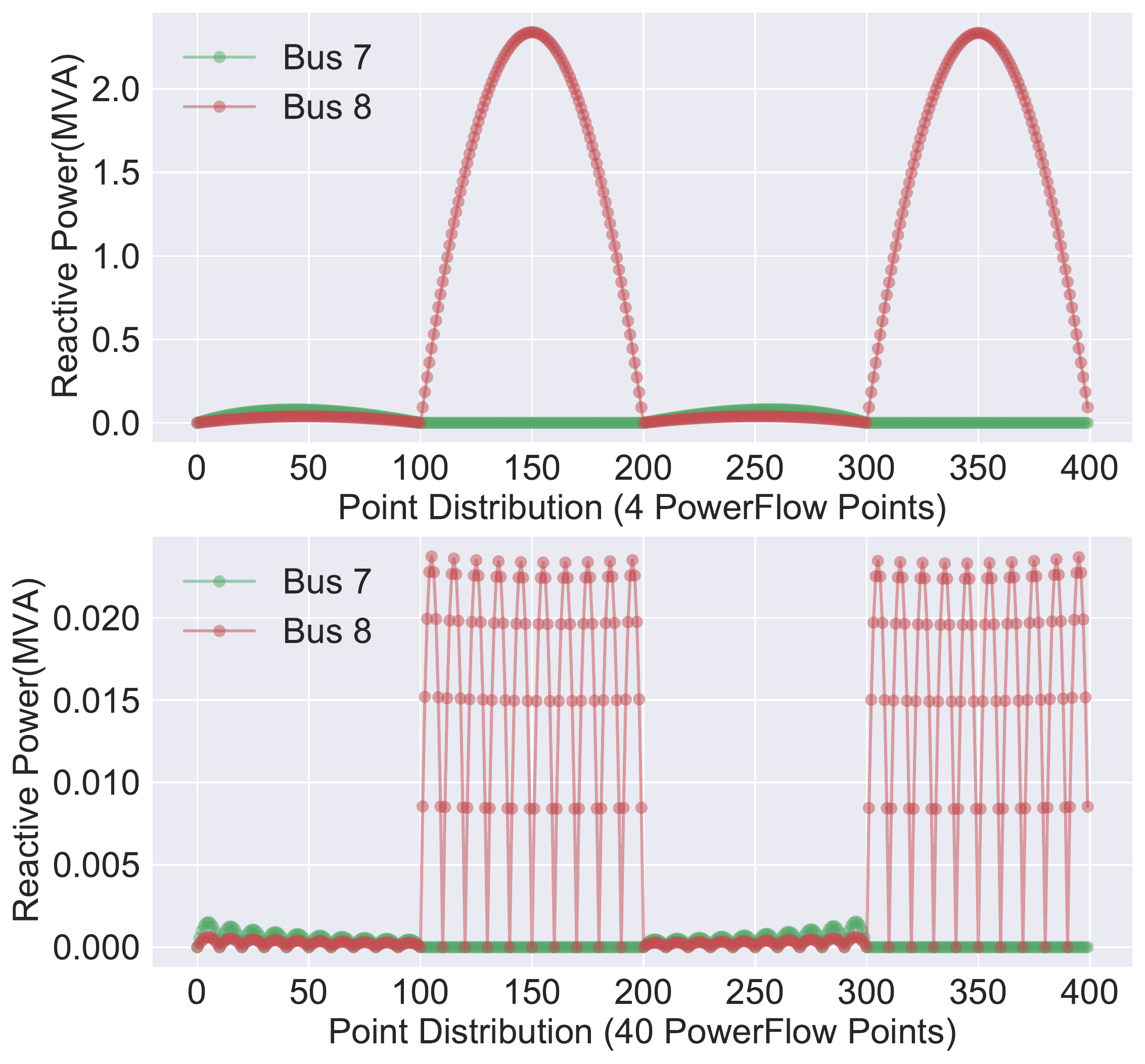}
         \caption{Zoomed hosting capacity region}
         \label{fig:Interpolation_Compare}
     \end{subfigure}
        \caption{Hosting capacity region correction}
\end{figure}

As shown in Table~\ref{tab:interpolation_time}, based on such four vertices, we compare three scenarios sharing the same boundary point number of 400. The first scenario derives all interpolations still based on powerflow calculation. The last one totally uses the proposed interpolation method based on known vertices. The middle one just compensates extra 9 points using powerflow calculation over each edge, and later utilizes the proposed method to get left points. Through comparing computation time cost in the same testing platform, the interpolation method is confirmed faster-speed than powerflow calculation. We also cite the results of exact region boundary points in Section.~\ref{subsubsec:case_2D_linear}, and compare them with interpolated points in the middle scenario in Table~\ref{tab:interpolation_time}. The good matching between these two points groups in Fi.~\ref{fig:Interpolation_Overall} confirm the validity of the proposed interpolation method.

The reason why we still adopt the middle scenario instead of the last one, even in higher computation time cost, it ensures limited reactive power correction volume. As shown in Fig.~\ref{fig:Interpolation_Compare}, during intepolation, the middle scenario owns maximal reactive power correction volume lower than 0.03MVA, while the worst scenario with only four powerflow points owns over 2MVA reactive power correction. It is still open for discussions on how to rationally select a proper powerflow point density, achieving a trade-off between computation time cost and reactive power correction volume in this case.

\begin{table}
    \centering
    \begin{tabular}{ c c c c}
    \toprule
         Total Points & PowerFlow Point & Interpolation Point & Computation Time(s)\\
       \midrule
        400 & 400 & 0 & 2.467\\
        400 & 40 & 360 & 1.526\\
        400 & 4 & 396 & 1.307\\
        \bottomrule
    \end{tabular}
    \caption{Table of vertices in hosting capacity region}
    \label{tab:interpolation_time}
\end{table}

\section{Hosting Capacity Region Reshaping}
The hosting capacity region exploitation technology has been deeply discussed in the previous section. We have payed lots of efforts to derive a hosting capacity region, and it may satisfy the DSO by providing an essential reference to help set respective capacity limits on each relevant POC. Meanwhile, the hosting capacity seems potential to create more values than that.

As mentioned above, in practices, although DSO sets relevant connection capacity limits to POCs, some users still need a larger connection capacity even with the willingness of higher connection expenses. Instead of physically upgrading grid cables, energy storage equipment installation is a promising alternative approach. With energy storage equipment integrated, the relevant hosting capacity region is equivalently reshaped. How to utilize such reshaping to help solve EES deployment problems is mainly discussed in this chapter.

\subsection{Minimal Energy Storage Capacity Quantification}
When ESS is integrated to a certain POC, its impacts on original hosting capacity region depends on its own output characteristics. Intended to quantify minimal ESS capacity to ensure a operation point, which was originally infeasible, become feasible again, we need to clarify the ESS characteristics in advance.

\subsubsection{Considering apparent power constraint}
Benefiting from a bidirectional inverter interface, the ESS can be flexible in its output power factor. In this scenario, we aim to calculate the minimal distance between the hosting capacity region and the operation point, which eventually quantify the minimal ESS capacity. There are many mature numerical methods investigated to calculate the distance from a point to a polygon in a planar space. However, in the beginning, we still choose to explain it geometrically for deeper understanding.

\begin{figure}
    \centering
    \includegraphics[width=0.9\linewidth]{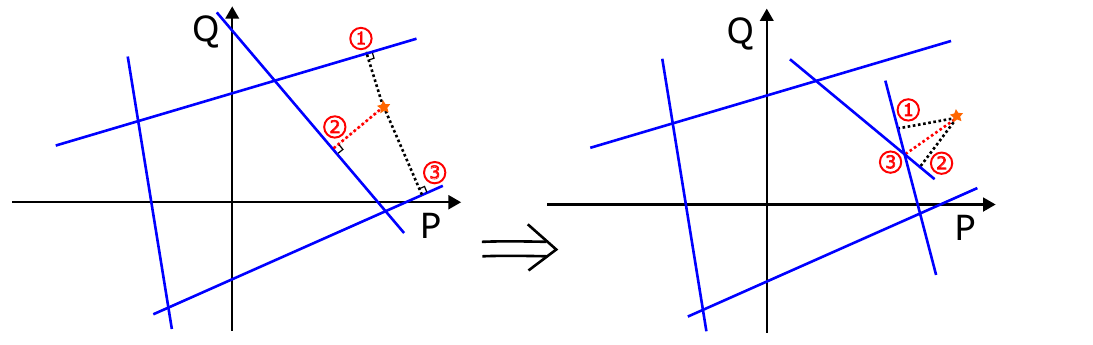}
    \caption{Minimal apparent power determination}
    \label{fig:Capacity_Region_Reshape_ESS_radius}
\end{figure}

As shown by the left picture in Fig.~\ref{fig:Capacity_Region_Reshape_ESS_radius}, when the polygon is not complicated and the operation point is well-posed, we can just calculate its distance to each boundary line as soon as this point meet this boundary. Unfortunately, as illustrated by the right picture, with polygon growing, the distance to one vertex can also be a solution, finally confusing and complicating our calculation progress.

\begin{figure}
    \centering
    \includegraphics[width=0.9\linewidth]{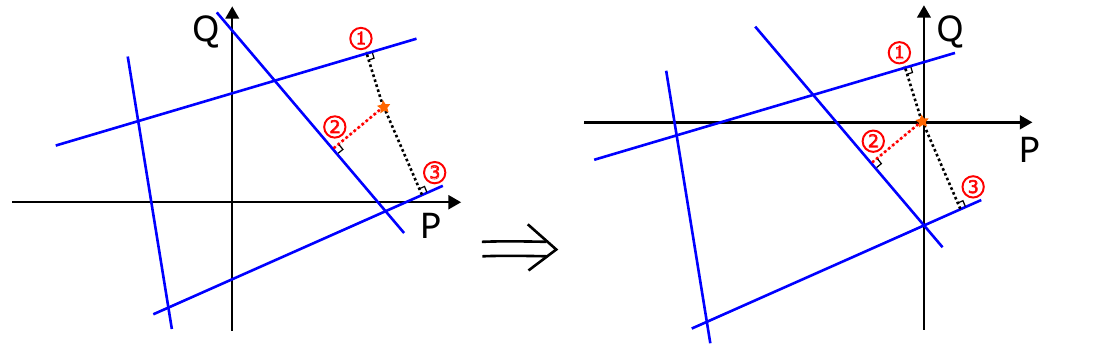}
    \caption{Coordinate shifting of hosting capacity region}
    \label{fig:Capacity_Region_Reshape_ESS_corrdinate}
\end{figure}

Meanwhile, from an algebraic perspective, this problem can be solved efficiently. Through coordinate shifting shown in Fig.~\ref{fig:Capacity_Region_Reshape_ESS_corrdinate}, this problem can be represented as
\begin{equation}
\label{eq:ESS_capacity_model}
    \textbf{min}~x^2+y^2~~~~~~~~~~
    \textbf{s.t. } (x,y)\text{ in shifted hosting capacity region}
\end{equation}

where $x,y$ are horizontal and vertical coordinate values in the shifted coordinate system.
Since the original hosting capacity region is defined by several linear bounds, the shifted one still keeps convex. The whole problem can be confirmed convex, which can be solved quickly while still ensuring the global optimum. The corresponding objective function value is equal to the expected minimal distance square. Such concise formulas are trying to answer the same questions about minimal EES capacity, while the grid model has been totally saved.

\subsubsection{Considering constant power factor}
\begin{figure}
    \centering
    \includegraphics[width=0.9\linewidth]{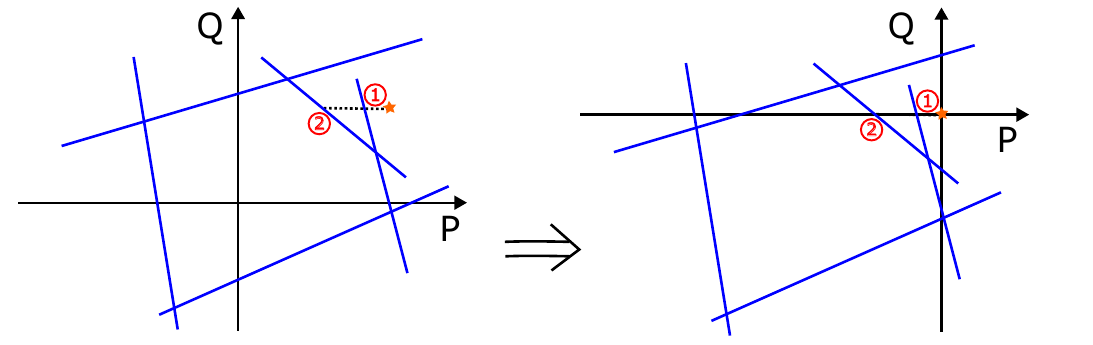}
    \caption{Coordinate shifting of hosting capacity region}
    \label{fig:Capacity_Region_Reshape_ESS_PF}
\end{figure}

In practices, ESS output power factor may be constrained constant, especially fot electrical-machine-based ESS, for instance flywheel applications. In such scenario, we can still use coordinate shifting method as shown in Fig.~\ref{fig:Capacity_Region_Reshape_ESS_PF}.If the power factor angle is set to $\alpha$, through substituting $y$ by tan($\alpha$)$x$, (\ref{eq:ESS_capacity_model}) is even simplified as there only exists only one variable. The final objective value can help quantify relevant ESS minimal capacity as well.

\subsection{General Optimal Storage Capacity Design}
In Section.~\ref{subsubsec:case_2D_linear}, we have assessed the hosting capacity region regarding pure active power over Bus~7 and Bus~8. If we want to install two ESS over these two POCs separately, we still need to figure out an optimal deployment scheme. Caused by geographical differences, there may be various installation expenses for ESS. If such installation fees still keep linear to it capacity, while Bus~7 and Bus~8 just shares various sloping rate $\beta$ and $\gamma$. Following the same principle in (\ref{eq:ESS_capacity_model}), the planning problem can be represented as

\begin{equation}
\label{eq:ESS_optimization_model}
    \textbf{min} \beta |x|+\gamma |y|~~~~~~~~~~
    \textbf{s.t. } (x,y)\text{ in shifted hosting capacity region}
\end{equation} 
The whole problem can be decomposed into four convex problems, where the symbol of $x$ and $y$ is fixed in each scenario.
Meanwhile, there exists another path to solve (\ref{eq:ESS_optimization_model}). Through equivalently converting (\ref{eq:ESS_optimization_model}) as below, we import two non-negative auxiliary variables $m$ and $n$ while keeping the same optimal objective function value.

\begin{subequations}
\begin{align}
    \textbf{min}~&\beta m+\gamma n\\
    \textbf{s.t.}~&m,n\geq0 \\
    & -m \leq x\leq m \\
    & -n \leq y\leq n \\
    & (x,y)\text{ in shifted hosting capacity region}
\end{align}
\label{eq:ESS_optimization_model_new}
\end{subequations}

\subsection{Case Study}
\begin{figure}
     \centering
     \begin{subfigure}[b]{0.472\textwidth}
         \centering
         \includegraphics[width=\textwidth]{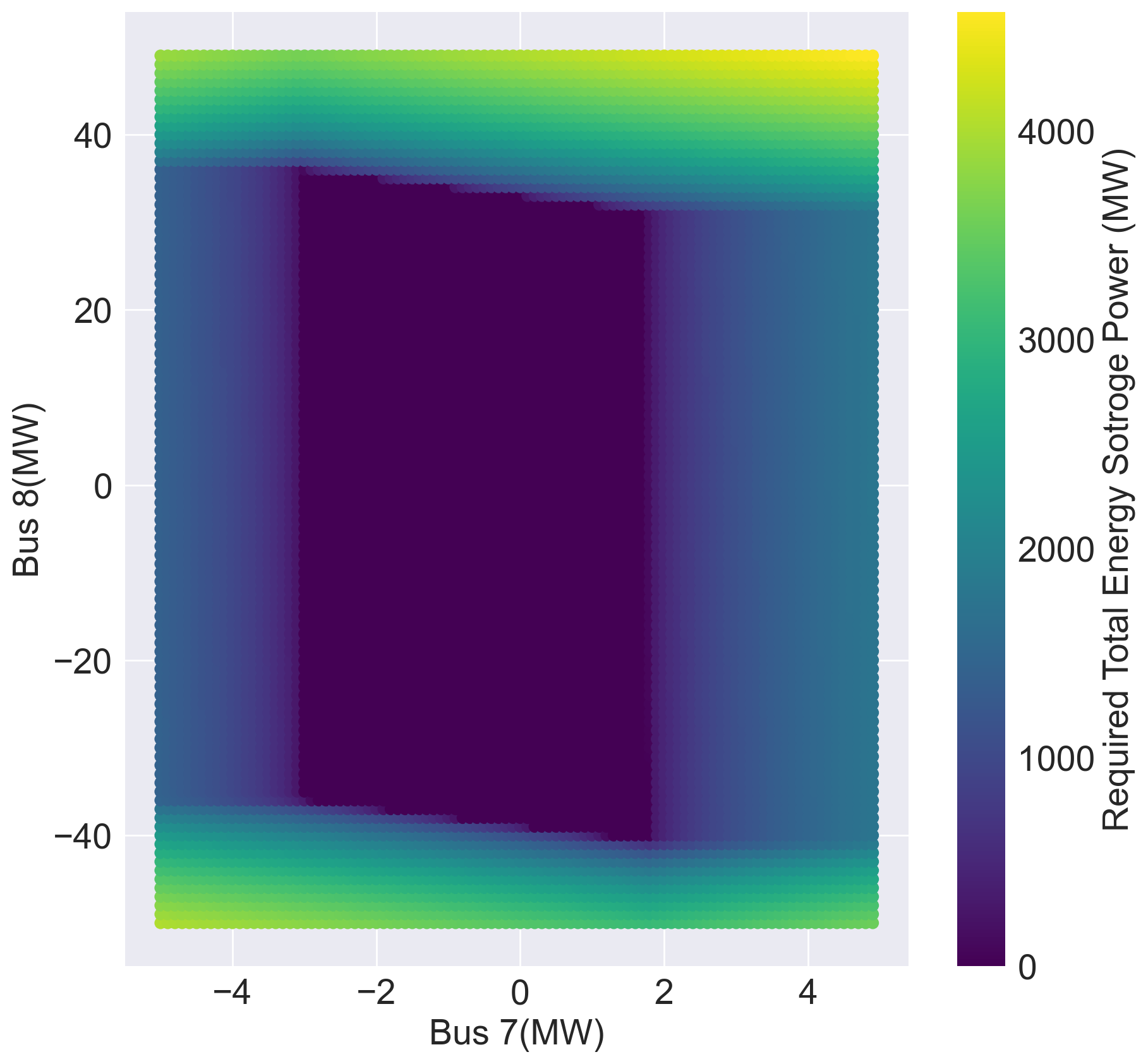}
         \caption{Optimal total energy storage capacity}
         \label{fig:Optimial_Capacity}
     \end{subfigure}
     \hfill
     \begin{subfigure}[b]{0.488\textwidth}
         \centering
         \includegraphics[width=\textwidth]{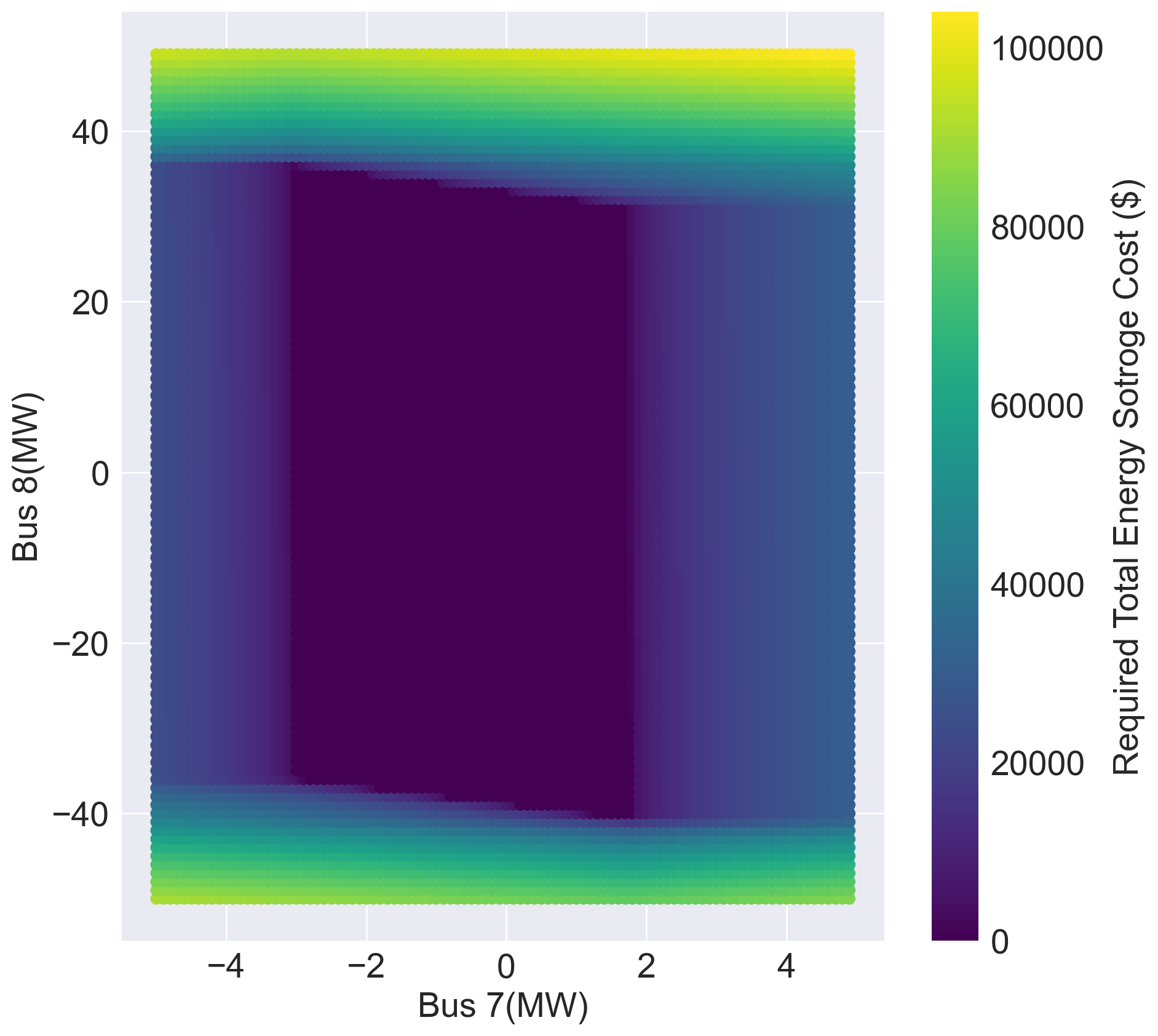}
         \caption{Optimal total energy storage cost}
         \label{fig:Optimal_Expense}
     \end{subfigure}
        \caption{Energy storage capacity and cost optimization}
\end{figure}

Similar to case studies in Section.~\ref{subsubsec:case_interpolation}, we still use four vertices to represent the feasible regions. Each edge of such quadrilateral can be taken as a linear constraint, defining a feasible half space by a cutting line. Looking into the storage capacity minimization problem, we can using the proposed algebraic optimization method to derive a conclusion. In order to verify the validity of such method, we have tested all operation points, where Bus~7 active power is from -5MW to 5MW and Bus~8 is from -50MW to 50MW. The corresponding optimal total energy storage capacity is denoted by color distribution Fig.~\ref{fig:Optimial_Capacity}. The dark blue part is equal to the feasible region, where the objective function value is equal to 0, indicating those operation points need no extra energy storage regulation.

Moreover, assuming the operation period set to 1 hour, based on battery cost data in \cite{cole2021cost}, we allocate 300\$/kWh for Bus~7 and 650\$/kWh for Bus~8. A general energy storage cost optimization problem is discussed. As illustrated in Fig.~\ref{fig:Optimal_Expense}, the cost distribution in various operation points has been provided. It can be seen from the results, that the dark blue region keeps equivalent to the original hosting capacity region.

Both testing cases above have confirmed the validity of the proposed optimization method, which is mainly based on the derived hosting capacity region. Answering the same question, unlike conventional scenario-based testing methods, our proposed one is independent from the original grid model, indicating lower memory requirement and higher computation efficiency.

\section{Summary}
Based on the term of "feasible region", this chapter extends the concept of "hosting capacity". Through converting the grid model into a more compact one, "hosting capacity region" is confirmed promising to further exploit the grid potential for power delivery. Such evolved concept also benefits energy storage deployment investigation in a more compact format, which is independent from the conventional grid model. Respective region assess schemes are exploited as well, whose validity has been confirmed with corresponding case studies.
In future, the authors will put more efforts into the high-dimensional assessment technology and more accurate capacity region correction methods. 

\printbibliography

@article{bollenintegration,
  title={Integration of Distributed Generation in the Power System},
  year={2011},
  author={Bollen, Math and Hassan, Fainan},
  publisher={Wiley Online Library}
}

@article{baran1989network,
  title={Network reconfiguration in distribution systems for loss reduction and load balancing},
  author={Baran, Mesut E and Wu, Felix F},
  journal={IEEE Power Engineering Review},
  volume={9},
  number={4},
  pages={101--102},
  year={1989},
  publisher={IEEE}
}

@article{low2014convex,
  title={Convex relaxation of optimal power flow—Part II: Exactness},
  author={Low, Steven H},
  journal={IEEE Transactions on Control of Network Systems},
  volume={1},
  number={2},
  pages={177--189},
  year={2014},
  publisher={IEEE}
}

@techreport{cole2021cost,
  title={Cost projections for utility-scale battery storage: 2021 update},
  author={Cole, Wesley and Frazier, A Will and Augustine, Chad},
  year={2021},
  institution={National Renewable Energy Lab. (NREL), Golden, CO (United States)}
}

@article{avis1997good,
  title={How good are convex hull algorithms?},
  author={Avis, David and Bremner, David and Seidel, Raimund},
  journal={Computational Geometry},
  volume={7},
  number={5-6},
  pages={265--301},
  year={1997},
  publisher={Elsevier}
}
\end{document}